\documentclass{ws-ijmpd}
\usepackage[super,compress]{cite}
\begin{document}

\markboth{C. Burigana, R.~D. Davies, P. de Bernardis, et al.}
{Recent Developments in Astrophysical and Cosmological Exploitation of Microwave Surveys}

\catchline{}{}{}{}{}
%
%%%%%%%%%%%%%%%%%%%%%%%%%%%%%%%%%%%%%%%%%%%%%%%%%%%%%%%%%%%%%%%%%%%%

\def\lsim{\,\lower2truept\hbox{${< \atop\hbox{\raise4truept\hbox{$\sim$}}}$}\,}
\def\gsim{\,\lower2truept\hbox{${> \atop\hbox{\raise4truept\hbox{$\sim$}}}$}\,}
\def\be{\begin{equation}}
\def\ee{\end{equation}}

\title{RECENT DEVELOPMENTS IN ASTROPHYSICAL AND COSMOLOGICAL EXPLOITATION OF MICROWAVE 
SURVEYS\footnote{Based on talks presented at the Thirteenth Marcel Grossmann Meeting on General Relativity, Stockholm, July 2012.}}

\author{CARLO BURIGANA}

\address{INAF--IASF Bologna, Via Piero Gobetti 101, I--40129 Bologna, Italy
\footnote{Istituto Nazionale di Astrofisica --
Istituto di Astrofisica Spaziale e Fisica Cosmica di Bologna, Via Piero Gobetti 101, I--40129 Bologna, Italy}\\
Dipartimento di Fisica e Scienze della Terra, Universit\`a degli Studi di Ferrara,\\
Via Giuseppe Saragat 1, I--44122 Ferrara, Italy\\
burigana@iasfbo.inaf.it}

\author{RODNEY D. DAVIES}

\address{Jodrell Bank Centre for Astrophysics,
Alan Turing Building, School of Physics and Astronomy,\\
University of Manchester,
Oxford Road, Manchester, M13 9PL, UK\\
rdd@jb.man.ac.uk}

\author{PAOLO DE BERNARDIS}

\address{Dipartimento di Fisica, Universit\`a La Sapienza, P. le A. Moro 2, I--00185 Roma, Italy\\
paolo.debernardis@roma1.infn.it}

\author{JACQUES DELABROUILLE}

\address{APC, AstroParticule et Cosmologie, Universit\'e Paris Diderot, CNRS/IN2P3, CEA/lrfu, \\
Observatoire de Paris, Sorbonne Paris Cit\'e, 10, rue Alice Domon et L\'eonie Duquet,\\ 
75205 Paris Cedex 13, France\\
delabrouille@apc.univ-paris7.fr}

\author{FRANCESCO DE PAOLIS}

\address{Dipartimento di Matematica e Fisica ``Ennio De Giorgi'', Universit\`a del Salento\\ 
%Via Arnesano, CP 193, I-73100 Lecce, Italy\\ 
INFN (Istituto Nazionale di Fisica Nucleare), Sezione di Lecce\\
Via Arnesano, CP 193, I--73100 Lecce, Italy\\
Francesco.DePaolis@le.infn.it}

\author{MARIAN DOUSPIS}

\address{IAS, bat 121, Universit\'e Paris-Sud 11, Orsay, F--91400, France\\
marian.douspis@ias.u-psud.fr}

\author{RISHI KHATRI}

\address{Max Planck Institut f\"{u}r Astrophysik, Karl-Schwarzschild-Str. 1, 85741, Garching, Germany\\
khatri@mpa-garching.mpg.de}

\author{GUO CHIN LIU}

\address{Department of Physics, Tamkang University, Tamsui District, New Taipei City, 251, Taiwan\\
liugc@mail.tku.edu.tw}

\author{MICHELE MARIS}

\address{INAF - Osservatorio Astronomico di Trieste, Via G.B. Tiepolo 11, Trieste, Italy\\
maris@oats.inaf.it}

\author{SILVIA MASI}

\address{Dipartimento di Fisica, Universit\`a La Sapienza, P. le A. Moro 2, I--00185 Roma, Italy\\
silvia.masi@roma1.infn.it}

\author{ANIELLO MENNELLA}

\address{Dipartimento di Fisica, Universit\`a degli Studi di Milano, Via Celoria, 16, Milano, Italy\\
INAF/IASF Milano, Via E. Bassini 15, Milano, Italy\\
aniello.mennella@fisica.unimi.it}

\author{PAOLO NATOLI}

\address{Dipartimento di Fisica e Scienze della Terra, Universit\`a degli Studi di Ferrara,\\
Via Giuseppe Saragat 1, I-44122 Ferrara, Italy\\
INAF--IASF Bologna, Via Piero Gobetti 101, I-40129 Bologna, Italy\\
paolo.natoli@gmail.com}

\author{HANS ULRIK NORGAARD-NIELSEN}

\address{DTU Space, Elektrovej, DK - 2800 Kgs. Lyngby, Denmark\\
hunn@space.dtu.dk}

\author{ETIENNE POINTECOUTEAU}

\address{CNRS, IRAP, 9 Av. colonel Roche, BP 44346, F-31028 Toulouse cedex 4, France\\
Universit\'e deToulouse, UPS-OMP, IRAP, F-31028 Toulouse cedex 4, France\\
etienne.pointecouteau@irap.omp.eu}

\author{YOEL REPHAELI}

\address{School of Physics, Tel Aviv University, Tel Aviv 69978, Israel\\
CASS, University of California, San Diego, La Jolla, CA 92093, USA\\
yoelr@wise.tau.ac.il}

\author{LUIGI TOFFOLATTI}

\address{Departamento de F\'\i{sica}, Universidad de Oviedo, c. Calvo Sotelo s/n, 33007 Oviedo, Spain\\
IFCA-CSIC, Instituto de F\'\i{sica} de Cantabria, avda. Los Castros s/n, 39005 Santander, Spain\\
ltoffolatti@uniovi.es}

\maketitle

\begin{history}
\received{18 January 2013}
\revised{18 January 2013}
\end{history}

\begin{abstract}
%The abstract should summarize the context, content
%and conclusions of the paper in less than 200 words. It should
%not contain any references or displayed equations. Typeset the
%abstract in 8 pt Times roman with baselineskip of 10~pt, making
%an indentation of 1.5 pica on the left and right margins.
In this article we focus on the astrophysical results and the related cosmological implications derived from recent microwave surveys, with emphasis to those coming from the {\it Planck} mission.
We critically discuss the impact of systematics effects and the role of methods to separate the cosmic microwave background signal from the astrophysical emissions and each different astrophysical component from the others. We then review of the state of the art in diffuse emissions, extragalactic sources, cosmic infrared background, and galaxy clusters, addressing the information they provide to our global view of the cosmic structure evolution and for some crucial physical parameters, as the neutrino mass. Finally, we present three different kinds of scientific perspectives for fundamental physics and cosmology offered by the analysis of on-going and future cosmic microwave background projects at different angular scales dedicated to anisotropies in total intensity and polarization and to absolute temperature.
\end{abstract}


\ccode{PACS numbers: 98.80.-k, 98.70.Vc, 98.65.-r, 98.54.-h, 98.54.Kt, 98.35.-a, 96.50.Dj.}

%\tableofcontents

\section{Introduction}

Since its discovery, the cosmic microwave background (CMB) represents a crucial probe for our general view of the Universe and the understanding of key aspects in cosmology and fundamental physics. Furthermore, microwave surveys are becoming more and more relevant for the comprehension of the physical and evolutionary properties of 
astrophysical structures at different cosmic epochs, from galactic to cosmological scales. 
Following the very important results from balloon-borne experiments, the NASA COBE and WMAP satellites, and recent ground-based projects, covering together a wide multipole range,
the available and forthcoming data products from the {\it Planck} mission\footnote{{\it Planck} is a project of the European Space Agency - ESA - with instruments provided by
two scientific Consortia funded by ESA member states (in particular the lead countries: France and Italy) with
contributions from NASA (USA), and telescope reflectors provided in a collaboration between ESA and a scientific
Consortium led and funded by Denmark.} will have a strong impact in these fields in the coming decades\footnote{This paper is based largely on the {\it Planck} Early Release Compact Source
Catalogue and publicly available publications by ESA and the   {\it Planck}
Collaboration, for what concerns the related aspects. Any material
presented here that is not already described in   {\it Planck}
Collaboration papers represents the views of the authors and not
necessarily those of the {\it Planck} Collaboration.}.
{\it Planck} instruments are in fact the most sensitive microwave receivers ever launched in space. 
Their sensitivity calls for a comparable level of systematic effect control, one of the main drivers in satellite and instrument design and currently key in data reduction and interpretation\cite{planck2011-1.4,planck2011-1.5,planck2011-1.6,planck2011-1.7}\,, a topic addressed in Section \ref{syst}.
Similarly, for high sensitive microwave observations, the accuracy in the recovery of the CMB properties largely relies on the capability to disentangle the cosmological signal from the astrophysical emissions
(the so-called foregrounds), as discussed in Section \ref{compsep}. Waiting for the {\it Planck} cosmological results, we focus here on currently available astrophysical discoveries based on the first {\it Planck} surveys possibly complemented by other sets of observations carried out at similar wavelengths and combined with surveys in other frequency domains, in a multifrequency approach. Section \ref{diffuse} is devoted to the diffuse emissions coming from the Solar System and the Galaxy, on the physical processes operating at various Galactic scales, and on their manifestation at microwave wavelengths.
In Section \ref{extra} we discuss the main properties derived on extragalactic sources at different cosmic distances, including the main evidencies regarding the cosmic infrared background.
Section \ref{clusters} is dedicated to the recent observational results on galaxy clusters and on their scaling relations, with the wealth of information they provide on baryon physics and their use as a probe for neutrino mass estimation.
Finally, in Section \ref{cmb} we review three different topics in CMB cosmology, possibly linked to fundamental physics, 
that will be addressed respectively by the forthcoming results from the {\it Planck} mission, by future high resolution 
ground-based experiments, and by the next generation of CMB spectrum projects.

%%%%%%
% Systematic Effects

\section{Control, assessment and removal of systematic effects in {\it Planck}}
\label{syst}

{\it Planck} orbits around the L2 Lagrangian point and scans the sky spinning at 1\,rpm in almost great circles with its Gregorian dual-reflector telescope pointing at 85$^\circ$ from the 
  spin axis\cite{tauber2010b,planck2011-1.1}\,. In the telescope focal plane the microwave photons are collected by two wide-band receiver arrays spanning a frequency interval ranging from $\sim$30\,GHz to $\sim$857\,GHz. The Low Frequency Instrument (LFI), is a coherent differential array based on 20\,K InP HEMT\footnote{Indium Phosphide High Electron Mobility Transistor} amplifiers currently working in three bands centered at approximately 30, 44 and 70\,GHz\,\cite{bersanelli2010}\,. The High Frequency Instrument is an array of bolometers cooled to 0.1\,K operating at six frequency bands centered at 100, 143, 217, 353, 545 and 857\,GHz\,\cite{lamarre2010}\,. {\it Planck} (full width half maximum, FWHM) resolution ranges from $33.3'$ to $4.3'$ going from 30 GHz to 857 GHz, and its final sensitivity per (FWHM$^{2}$ resolution element is in the range of $\sim 2-14$ $\mu$K$/$K in terms of $\delta T / T $ for frequencies $\nu \le 353$GHz.
The life of {\it Planck} largely exceeded the early plan.
Five all-sky surveys has been accumulated with HFI, while LFI is planned to operate up to about the end of Summer 2013, so completing eight all-sky surveys.

  In Table~\ref{tab_syseffect_list} we list the main systematic effects in {\it Planck} according to their source and provide few notes about their control and residual impact on science.
  
  \begin{table}[h!]
    \tbl{Main systematic effects in {\it Planck}}
    {\begin{tabular}{l l p{6.5cm}}
      \hline
	 \textbf{Category} & \textbf{Effect} & \textbf{Notes}\\
	 \hline
	 \textbf{\textit{Optics}}\cite{sandri2010,maffei2010} & &\\
	& Side-lobes\dotfill & Galaxy and CMB dipole pickup by main and sub-reflector spillovers. Negligible effect on temperature maps, needs to be removed at low frequency for polarization analysis. \\
	 \textbf{\textit{Detectors}} & &\\
	 & Cosmic ray hits\dotfill & Affect bolometric detectors. Removed from timelines via template fitting. \\
	 & 1/$f$ noise\dotfill & Affects radiometric and bolometric detectors. In the LFI the 1/$f$ contribution is limited to max 3\% by differential measurement strategy and de-striping 
	 algorithms\cite{kurki-suonio2009}\,.\\
	 & Bandpass mismatch\cite{leahy2010}\,\dotfill & Affects primarily radiometric detectors. Negligible impact on temperature. Corrected in polarisation at map level exploiting polarized source measurements (Crab Nebula).\\
	 \textbf{\textit{Electronics}} & &\\
	 & 1-Hz spikes\dotfill & Affects LFI data. Removed from timelines by template fitting.\\
	 \textbf{\textit{Thermal}}$^\dagger$ & &\\
	 & 300\,K fluctuations\dotfill & In principle affect both instruments. Inherent hardware stability is compliant with scientific requirements. \\
	 & 20\,K fluctuations\dotfill & Affect mainly LFI. Inherent hardware stability is compliant with scientific requirements.\\
	 & 4\,K fluctuations\dotfill & Affect both instruments. Inherent hardware stability is compliant with scientific requirements.\\
	\hline
	\multicolumn{3}{p{12cm}}{$^\dagger$In the LFI a combination of the differential measurement strategy with calibration and de-striping further reduce the effect.}
      \end{tabular}\label{tab_syseffect_list}}
  \end{table}
  
\begin{figure}[h!]
    \centerline{\psfig{file=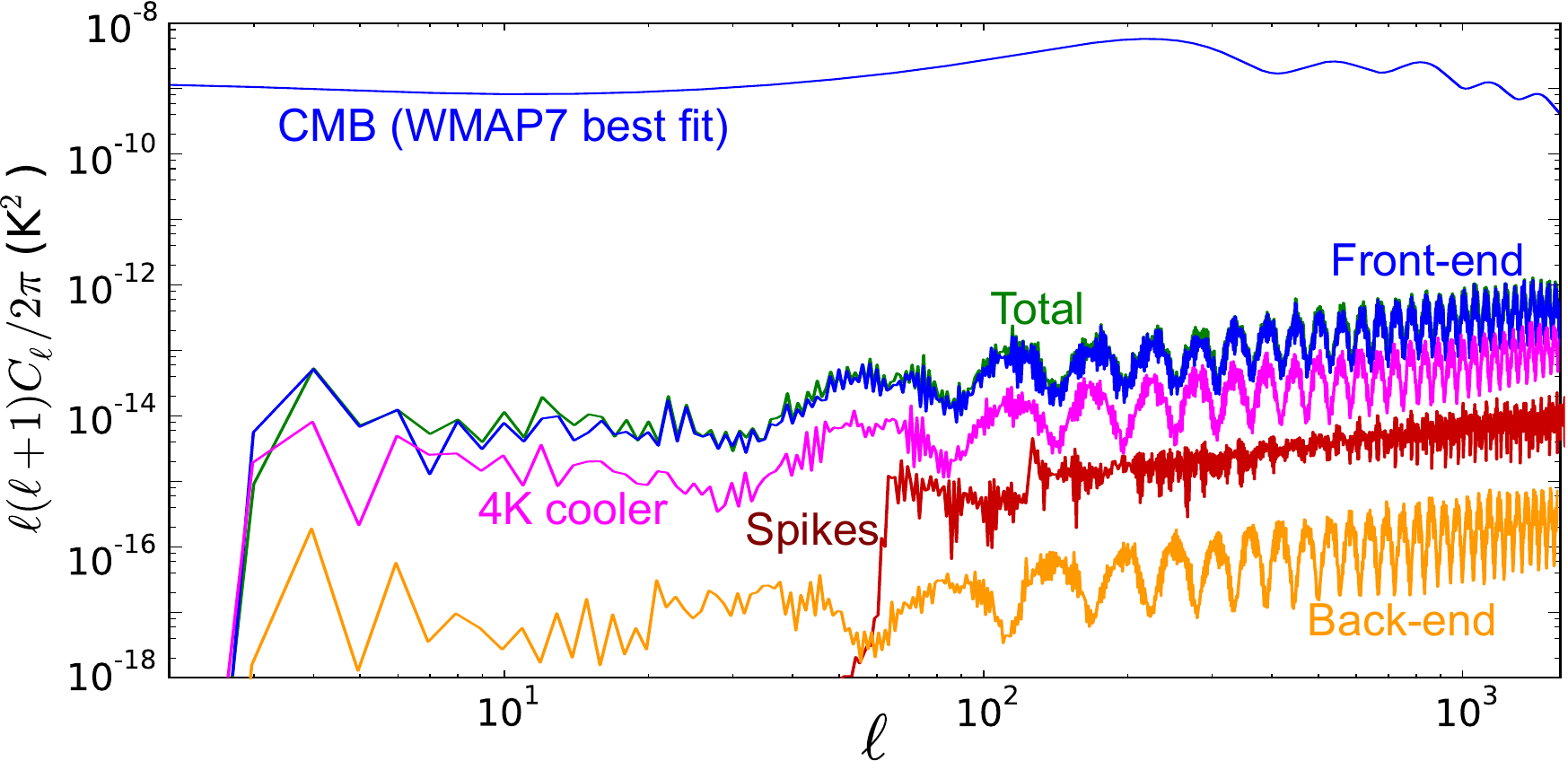,width=8cm}}
    \caption{Impact of main systematic effects in {\it Planck} LFI on temperature angular power spectrum (APS) as function of the multipole $\ell$.\label{fig_syseffect_power_spectra}}
  \end{figure}

  In the LFI we have generated timelines, maps and power spectra of thermal effects and 1-Hz spikes using in-flight scientific and housekeeping data coupled with transfer functions measured during ground tests. Table~\ref{tab_syseffect_budget_maps} reports the peak-to-peak and rms effect on full-sky temperature maps of these effects, while in Fig.~\ref{fig_syseffect_power_spectra} we show their expected temperature angular power spectra after component separation. The effect of component separation (see next section) has been reproduced by mixing the systematic effects maps using the same mixing matrix used to extract the {\it Planck} CMB map (not reported in this paper) from the individual frequency maps. 
  
  Our current analysis confirms that the level of systematic effects rejection is in line with pre-launch expectations and will allow full exploitation of the science encoded in the CMB signal.
  
  \begin{table}[h!]
    \tbl{Effect on {\it Planck} LFI maps of the main systematic effects}
    {\begin{tabular}{l c c c c c c c c}
      \hline
	& \multicolumn{2}{c}{30\,GHz} & & \multicolumn{2}{c}{44\,GHz} & & \multicolumn{2}{c}{70\,GHz}\\
	& \multicolumn{2}{c}{$[\mu\mathrm{K}]$} & & \multicolumn{2}{c}{$[\mu\mathrm{K}]$} & & \multicolumn{2}{c}{$[\mu\mathrm{K}]$} \\
	\cline{2-3}\cline{5-6}\cline{8-9}\\
	\multicolumn{1}{c}{Channel} & p-p & rms & & p-p & rms & & p-p & rms\\
	\hline
	1-Hz spikes\dotfill & 4.00 & 0.45 & & 1.51 & 0.15 & & 2.56 & 0.30\\
	Thermal fluct\dotfill & & & & & & & & \\
	\hspace{0.5cm}Back-end\dotfill & 1.27 & 0.11 & & 0.63 & 0.05 & & 2.70 & 0.24 \\
	\hspace{0.5cm}Front-end\dotfill & 1.05 & 0.23 & & 1.15 & 0.22 & & 1.12 & 0.21 \\
	\hspace{0.5cm}4\,K loads\dotfill & 9.76 & 0.98 & & 9.73 & 0.98 & & 1.30 & 0.16 \\
	Total\dotfill & 10.92 & 1.10 & & 9.73 & 0.98 & & 4.28 & 0.45 \\
	\hline
      \end{tabular}\label{tab_syseffect_budget_maps}}
  \end{table}

%%%%%%%%%
% Comp. sep. 

\newcommand{\va}{\mathbf{a}}
\newcommand{\vn}{\mathbf{n}}
\newcommand{\vs}{\mathbf{s}}
\newcommand{\vw}{\mathbf{w}}
\newcommand{\vx}{\mathbf{x}}
\newcommand{\tA}{\mathsf{A}}
\newcommand{\tR}{\mathsf{R}}
\newcommand{\tW}{\mathsf{W}}
\newcommand{\Id}{\mathsf{Id}}

\section{Component separation}
\label{compsep}

As widely discussed in the next section, the sky emission, at a given frequency, is a superposition of emission from various sources.
%One can distinguish them by physical process (thermal, synchrotron, Bremsstrahlung, etc.), and by site or physical medium of emission: 
%the interstellar medium (ISM) in our own Milky Way or in external galaxies; the original plasma emitting the primary CMB; 
%hot electron gas in clusters of galaxies, generating thermal Sunyaev-Zel'dovich emission, etc.
The plausible contamination of the observable primary CMB by foreground emission has always been a source of concern for CMB observations. However, the level of foreground contamination, at high Galactic latitude and at frequencies between 50 and 200 GHz, is low enough that for $\ell$ less than about 2500 the temperature (or total intensity) APS, $C_\ell^{TT}$, can be accurately measured with only minor masking of the regions most contaminated by foregrounds (Galactic interstellar medium (ISM) and bright extragalactic compact sources). On smaller scales, emission from a background of blended faint extragalactic sources contributes a significant fraction of the observed power\cite{2003ApJS..148...97B,2011ApJ...739...52D,2010ApJ...718..632H}\,.

For a sensitive mission such as {\it Planck}, foreground emission, rather than instrumental noise, sets the limit of the accuracy of the measurement of the CMB APS. This limit depends on the effectiveness of any foreground-cleaning technique used to separate the primary CMB emission from foregrounds. Hence, the development, comparison, and optimization of component separation methods has been an important activity in the {\it Planck} Collaboration during the preparation of the 
%mission\cite{1999NewA....4..443B,2008A&A...491..597L,2009LNP...665..159D,2012arXiv1210.1416M}\,.
mission\cite{1999NewA....4..443B,2008A&A...491..597L,2009LNP...665..159D,2012A&A...548A..51M}\,.

\subsection{Modeling sky emission}

%Whether it is possible to identify, in a set of observed sky maps, the individual contribution of each source of interest, is tightly connected to a multicomponent model of sky emission that both serves as a 
%framework for interpreting the observations, and as a summary of our knowledge about the astrophysical emitters. 
A multicomponent model of sky emission serves both as a framework for analyzing and interpreting the observations, and as a summary of our knowledge about the astrophysical emitters. Recently, such a model of sky emission, the {\it Planck} Sky Model (PSM), has been put together for this purpose\cite{2012arXiv1207.3675D}\,. It is based on an underlying $\Lambda$CDM cosmological model with associated standard parameters\footnote{The spectral index and amplitude of scalar perturbation, $n_s$ and $A_s$, the ratio between the amplitude of tensor and scalar perturbations,
$r \! = \! A_t/A_s$, the density parameters of matter, baryons, dark energy, $\Omega_{\rm m}$, $\Omega_{\rm b}$, $\Omega_{\Lambda}$, the Hubble constant,
$H_0$, etc., the default values of which are set to the current best fit.}. In this framework, the emission of the CMB, of galaxy clusters, and of high-redshift galaxies in haloes of large-scale density contrast, is described on the basis of their statistical properties (angular power spectra of the CMB, cluster number counts, source number counts and halo occupation distribution as a function of luminosity, of redshift, and of spectral energy distribution). Known bright radio and infrared sources are modeled using  extrapolations of their measured fluxes at various frequencies. 
Galactic diffuse emission 
%from the Galactic ISM 
is modeled as a mixture of several components: synchrotron radiation from energetic electrons spiraling in the Galactic magnetic field, thermal dust emission, free-free (Bremsstrahlung) emission from the warm interstellar medium, the so-called `anomalous' dust emission from small spinning dust grains. Molecular line emission could also contribute to the signal in some frequency bands, that are typically relatively wide in order to increase the measurement sensitivities. In particular, the emission from the CO line is clearly seen in {\it Planck} data sets that allow to produce all-sky accurate maps of 
this signal\cite{planck2011-1.7}\,. 
%data sets\cite{2011A&A...536A...6P}\,.  

\subsection{Basics of component separation}

Consider a single pixel $p$ in a set of maps observed at various frequencies, indexed by $\nu$. The signal observed at frequency $\nu$, in pixel $p$, can be written as
\begin{eqnarray}
x(\nu,p) = \sum_i a_i(\nu,p) \, s_i(p) + n(\nu,p),
\end{eqnarray}
or, in vector-matrix format
\begin{eqnarray}
\vx(p) = \tA(p) \, \vs(p) + \vn(p).
\label{eq:linearmix}
\end{eqnarray}
If we know the frequency dependence $\tA(p)$ of each foreground in pixel $p$, our problem is just to invert a (set of) linear system(s) to find the reference component templates $\vs(p)$. In the limit where the instrumental noise $\vn(p)$ is small, the inversion is implemented using the inverse of $\tA$ (or, for non-square systems, the left pseudo-inverse $\tA^{-1}_{\rm left} \! = \! [\tA^t \tA]^{-1}\tA^t$). Otherwise, classical linear solutions such as least square (LS) or Wiener inversion can be used. They require, however, the prior knowledge of the covariance matrix $\tR_n(p)$ of the instrumental noise in each pixel, and possibly also of that of the signal, $\tR_s(p)$. The LS solution is 
\begin{eqnarray}
\widehat{\vs}_{\rm LS}(p) = \left [ \tA^t \tR_n^{-1} \tA \right ]^{-1}  \tA^t \tR_n^{-1} \vx(p),
\end{eqnarray}
and the Wiener one is
\begin{eqnarray}
\widehat{\vs}_{\rm Wiener}(p) = \left [ \tA^t \tR_n^{-1} \tA + \tR_s^{-1} \right ]^{-1}  \tA^t \tR_n^{-1} \vx(p),
\end{eqnarray}
where $\tA^t$ denotes the transpose of $\tA$. Most of the time however, neither the `mixing matrix' $\tA$, nor the statistical properties of the signal, are known. Sometimes even the noise covariance is not well known. One must then find a way to estimate them from the data themselves prior to inverting the linear mixture and recovering an estimate of each of the components.

\subsection{Blind component separation}

Blind separation of linear mixtures is a classical field of research in signal and image processing. 
%Applications range from audio processing to  medical imaging, astronomy, and more. 
Typically, the data model is of the form of Eq.~(\ref{eq:linearmix}), except that $\tA$ does not depend on $p$, and noise is often not an issue. The main problem is then to decide how many components exist in the data, and determine $\tA$ (or find a matrix $\tW$ that inverts the system, without explicitly estimating $\tA$). The main idea is to use the (assumed) statistical independence of the various components $\vs$. Blind component separation applied to CMB observations must address specific issues: error estimation, ill-conditioned covariance matrices, correlations between some of the components, spherical data sets. This has led the CMB community to adapt classical independent component analysis (ICA) methods for their analyses. 

The FastICA method\cite{2000MNRAS.318..769B,2002MNRAS.334...53M} aims at inverting the linear system using the matrix $\tW$ that maximizes a measure of the non-Gaussianity of the component maps. The method, however, is not very effective at distinguishing Gaussian CMB from Gaussian noise. It also fails to exploit the strong spatial correlation of the CMB and of most of the diffuse foreground emission.

Spectral matching ICA (SMICA)\cite{2003MNRAS.346.1089D,2005MNRAS.364.1185P,2008ISTSP...2..735C}\,, is a flexible method that maximizes the likelihood of a parametric model of $\tA$, $\tR_s$ and $\tR_s$ by minimizing the spectral mismatch between empirical and modeled second-order statistics of the observed maps. It is particularly useful for measuring a CMB APS, or parameters that model it, directly from multifrequency data. It has been used for predicting the errors on the tensor to scalar ratio $r$ that can be reached by future CMB B-modes experiments\cite{2009A&A...503..691B}\,. Very similar in spirit, although many implementation details vary, the correlated component analysis (CCA) method has been developed to deal specifically with correlated components\cite{2006MNRAS.373..271B}\,. Once second-order statistics of components and noise are obtained, they are used to invert the linear system.

Neural networks provide another attractive solution for finding either coefficients that invert the linear mixture, or only those coefficients that recover the CMB specifically\cite{2008Ap&SS.318..195N}\,. The method seems to perform well both on simulations\cite{2009AN....330..863N} and on real data\cite{2010A&A...520A..87N}\,. The impact of the training of the neural network, however, is hard to evaluate, and the propagation of errors not straightforward. These two limitations are serious for CMB data analysis, and may explain why the method has not received more attention so far.

A completely different point of view is taken in [\refcite{2006ApJ...641..665E}]. Instead of learning (explicitly or implicitly) the model (or part of it) in the data themselves, a parametric model of all relevant foregrounds is assumed a priori. Specifically, an amplitude and parameters defining an appropriate emission law are assigned to each emission process in each sky pixel. The value of all parameters are then found using a Monte-Carlo Markov chain (MCMC) algorithm. This is clearly the appropriate approach for measuring efficiently parameters of a known model, and is hence of much interest for component separation. It is also very flexible, as the foreground model can be chosen freely. The main caveat is that while in theory this method provides a complete likelihood (and hence errors) for all parameters, in practice the main uncertainty is whether the assumed parametric model is correct. Goodness of fit is not a fully satisfactory criterion: as well known, it is always possible to fit a limited data set with a wrong model, provided the number of parameters used in the fit is large enough. The method is hence of interest only when many different channels of observation are available. An 
alternative implementation of this idea has been proposed by [\refcite{2009MNRAS.392..216S}].

Finally, yet another approach, based on a different optimization criterion to recover the mixing matrix $\tA$, has been proposed by [\refcite{2008StMet...5..307B}]. It uses a likelihood penalization imposing a sparse representation of sky components in some over-complete dictionary of functions that serve as a redundant basis. The first implementation of the method, although promising and conceptually interesting, was not optimized for real-life data processing, due to lack of proper handling of different map resolutions and of the non-stationarity of the component emissions. This has been improved with a recent version that uses wavelet decompositions, allowing variations in both pixel and harmonic space of the linear combinations of the maps used for CMB recovery\cite{2012arXiv1206.1773B}\,. 
Such localization makes it possible to relax the restrictive condition that the emission of each component should be decomposable into the product of a pixel-independent emission law $A(\nu)$ and a spatial template $s(p)$.
%%, and has also been proposed and used earlier in the implementation of an approximation of the LS solution by \cite{2009A&A...493..835D,2011MNRAS.418..467R,2012MNRAS.419.1163B}. %%This is discussed in more detail in the next paragraph.

\subsection{The internal linear combination and variants}

Bypassing in some way the need to rely on a specific model of all foreground emission (especially rigid linear mixtures in which $\tA$ does not depend on $p$) is an appealing option. The frequency scaling of CMB anisotropies themselves being known to be the derivative with respect to temperature of a 2.725$\,$K blackbody, independently of $p$, one may write a simplified model of the observed maps as
\begin{eqnarray}
\vx(p) = \va \, s(p) + \vn(p),
\label{eq:cmbobs}
\end{eqnarray}
where $s(p)$ is the CMB map, $\va$ the CMB frequency scaling, and all the unknown (or poorly known) noise and foreground contamination is dumped together into a single noise term $\vn(p)$. The LS reconstruction of the CMB map is then
\begin{eqnarray}
\widehat{\vs}_{\rm LS}(p) = \frac {\va^t \tR_n^{-1} \vx(p)}{\va^t \tR_n^{-1} \va}.
\end{eqnarray}
This may seem impossible to implement without knowing $\tR_n$ (which now includes unknown foregrounds correlated between channels). However, under the hypothesis that $s(p)$ is not correlated with $\vn(p)$, the covariance $\tR_x$ of the observed maps is
%\begin{eqnarray}
$\tR_x = \sigma_s^2\va\va^t + \tR_n$,
%\end{eqnarray}
where $\sigma_s^2$ is the variance of the CMB map. It is then straightforward to show, using the Woodburry inversion formula, that $\va^t \tR_x^{-1} \propto \va^t \tR_n^{-1}$, and hence that the LS solution can be rewritten as
\begin{eqnarray}
\widehat{\vs}_{\rm LS}(p) = \frac {\va^t \tR_x^{-1} \vx(p)}{\va^t \tR_x^{-1} \va}.
\end{eqnarray}
This form is easily implemented using empirical estimates $\widehat \tR_x$ of $\tR_x$ obtained on the data themselves. This solution is also obtained as the constrained minimization problem of finding the linear combination $\vw^t\vx$ of the inputs that has minimum variance under the `CMB preserving' condition $\vw^t\va=1$. It is classically called the internal linear combination (ILC) method.

The ILC method, for its simplicity and robustness, has been used for the analysis of COBE-DMR, WMAP, and {\it Planck} data\cite{1992ApJ...396L...7B,2007ApJS..170..288H,planck2011-1.7}\,.
%2011A&A...536A...6P}\,.
Variants that compute weights in different regions of pixel, harmonic or wavelet space for CMB temperature or polarization have been 
derived\cite{2003PhRvD..68l3523T,2007ApJ...660..959P,2009A&A...493..835D,2012MNRAS.419.1163B}\,. Extensions for recovering more components than just the CMB are discussed 
by [\refcite{2011MNRAS.410.2481R,2011MNRAS.418..467R}].

It is important to note that the ILC is prone to subtle biases, which must be understood and controlled for scientific analyses based on ILC maps. The first bias, a loss of some modes of the original CMB and hence of CMB power, is due to empirical correlations between the CMB and the contaminants, and is discussed at length in the appendix of [\refcite{2009A&A...493..835D}]. The second is an amplification of calibration errors in the observed channels (or errors in the assumed frequency scaling of the component of interest), and is discussed in detail by [\refcite{2010MNRAS.401.1602D}].

\subsection{Error assessment and masking}

One of the most crucial questions, once the component separation is performed, is the assessment of errors. How well does a method perform? While it is easy to propagate errors in a fit, the problem in component separation is that modeling errors dominate the uncertainties. Nonetheless, three approaches can give an idea of component separation performance. 

First, one can compare the results obtained with methods that are conceptually very different. If, however, results are very method-dependent, 
%(as is usually the case), 
as usual,
then one must either explain the differences and discard the method(s) thought to be less effective, or, for the post-analysis of the output map(s), mask (or flag as plausibly contaminated) any sky region in which agreement cannot be achieved. 

Second, one may test methods on simulations that are as realistic as possible. This has been one of the original motivations for the development of the PSM. However, the performance of some component separation methods is very sensitive on very subtle details about the sky emission. The refinement of the model and the separation of components are thus two complementary parts of a global, iterative, data analysis chain. 

Finally there is a third method, that permits to identify regions of the sky where the number of channels available is not sufficient to separate all emissions. Consider noisy observations of unspecified sky signals, $\vx(p) = \vs(p) + \vn(p)$.
Usually, the sky and noise components are pairwise de-correlated, and thus $\tR_x = \tR_s + \tR_n$.
We now suppose that $\tR_n$ (instrumental noise only) is reasonably well known. Then we can whiten the observations (by multiplication by the square root of $\tR_n$). For the new data set, we have $\tR_{x} = \tR_{s} +  \Id$. In the basis of diagonalisation of $\tR_{x}$, the covariance becomes $\tR_{x} = \Delta + \Id$.
The number of (local) eigenvalues of $\tR_{x}$ significantly larger than unity is the dimension of the space spanned by measurable signal components. If all eigenvalues are larger than unity, then there are locally more independent astrophysical emissions than can be separated without external information. If, however, only some of the eigenvalues are significantly larger than unity, then in principle the data set is redundant enough for blind component separation, for instance with methods such as SMICA and the ILC which, implemented in wavelet space, are expected to perform very satisfactorily.

%%%%%%%%%%%%%%

\section{Diffuse foregrounds}
\label{diffuse}

Except for the averaged (monopole) signal, microwave and sub-mm surveys are dominated at large angular scales by the diffuse signals from the Solar System and the Milky Way, emerging as foreground sources with patterns particularly prominent close to the ecliptic and Galactic plane, respectively, and, typically, remarkable up to few tens of degrees from them. The observer position plays a significant role in the study of the former (while it can be fully neglected for the latter, at least at the angular resolutions relevant here). For this reason, 
a special care in the application of the component separation methods described in previous section, if not a fully different approach, is required in this case, 
since, in general, the time dependence of the signal should be accounted for.

%%%%%%%%%
% SS diffuse emissions

\subsection{Solar System diffuse emissions}

\def\Tdust{T_{\mathrm{dust}}}
\def\tauDust{\tau_{\mathrm{dust}}}
\def\almSS{a_{\ell,m,\mathrm{SS}}}

\def\Cl{C_\ell}
\def\ClCMB{C_\ell^{\mathrm{CMB}}}
\def\ClSS{C_\ell^{\mathrm{SS}}}

\begin{figure}[t]
\centering
\includegraphics[height=1.7in,width=3.in]{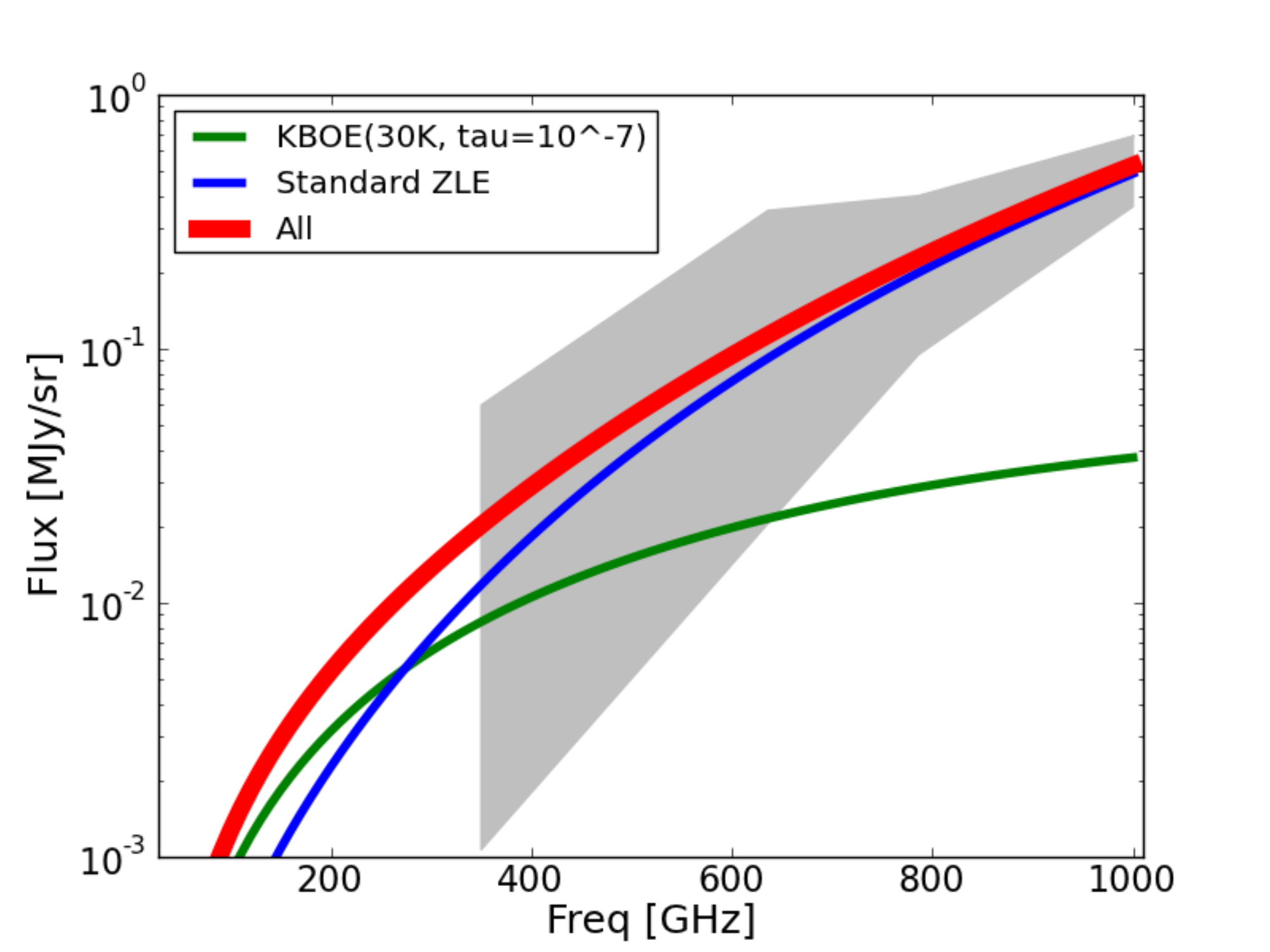}
 \caption{The expected spectral energy distribution (SED) for the sky averaged
ZLE (blue line) and an hypothetical cold dust component with $\Tdust=30$~K,
$\tau=10^{-7}$ (green line) and their sum (red line)
compared with the 
COBE/FIRAS spectrum
% provided by [\refcite{FD02}] 
(gray patch). }
\label{fig:zle}
\end{figure}

Solar System provides the most fore of all the foregrounds. 
In particular 
Zodiacal Light Emission (ZLE), i.e.
the emission from Interplanetary Dust Particles (IDPs), dominated
the sky signal at short wavelengths.
At wavelengths shorter than 12~$\mu\mathrm{m}$ ZLE
is mainly due to scattering of
solar radiation, while at longer wavelengths
thermal emission is the most important generation mechanism\cite{kelsalletal1998,FD02}\,.
ZLE is usually not accounted in CMB studies. In fact, since
the ZLE flux below 1~THz 
decreases with $\nu^4$ (see Fig.~\ref{fig:zle}, where the COBE/FIRAS spectrum is provided by [\refcite{FD02}]), 
i.e. as a modified blackbody with $\Tdust=240$~K,
its contribution would be significantly smaller than the others foreground signals at CMB related frequencies.

However Fig.~\ref{fig:zle} is based just on the observed emission of 
the IDPs at frequencies above 1~THz,
which is dominated by the population of grains between Earth and Jupiter orbits.
Given dust orbiting the Sun is removed in short times by Pointing-Robertson decay, 
radiation pressure, grains--grains collisions and planetary encounters a mechanism to continuously 
refurbish the population of IDPs is needed. In the inner Solar System the main 
contributors are comets and asteroids, even if 
some dust is supposed to come from interstellar space.
Other sources of dust are know to be effective far from the Jupiter orbit,
in particular the erosion and mutual collisions of trans-Neptunian objects (TNOs) and Kuiper Belt objects (KBOs) produced a second band of
dust in the outer Solar System, detected by deep space probes. 
We will denote with KBOE the 
%contribution to the Solar System 
diffuse 
%foreground 
emission from this class of IDPs.

Being at heliocentric distances much larger than IDPs responsible for standard
ZLE, particles responsible for KBOE would be quite cold having
$\Tdust \approx 30-60$~K or lower.
The optical depth below 1~THz 
is unknown. It depends on the balance between the mechanisms of dust production
either mutual collisions between larger bodies or erosion by interstellar dust,
the former tends to produce larger grains then the latter, the dust collision rate,
the composition of dust, the geometrical distribution of dust and the relative efficiency
of production and destruction mechanisms. However it hardly will exceeds the optical depth
for standard ZLE which is about $10^{-7}$. Two reasons make hard the detection of KBOE: $(i)$ 
at high frequencies the ZLE emission would overwhelm the KBOE, 
as evident from the 
example of SED in Fig.~\ref{fig:zle}; $(ii)$ the main method to separate the ZLE from background emission,
basically from Galactic dust, is to measure the seasonal dependence of 
ZLE signal for a fixed line--of--view, as the observer moves within the Solar System
while surveying the sky.
Such effect amounts to at most $7\%$ of the emission, more than 90\% of the modulation
comes from dust within 3~AU from the Earth, so that the seasonal dependence
from dust at heliocentric radii of 50~AU would be largely negligible.

It is evident how Solar System diffuse emission may act as a foreground for CMB in two ways.
The simplest case is when templates for Galactic and extragalactic emission
at frequencies relevant for CMB observations are extrapolated from 
maps obtained at wavelengths where the ZLE is an important source of systematics.
This is typical of CMB observations at frequencies  higher than 100~GHz,
where a template dust map is needed. 
Such maps are produced by extrapolating at CMB frequencies maps from observations
made at several tens of $\mu\mathrm{m}$ where the ZLE is very strong and must be removed.
In this case the ZLE will represent a sort of indirect foreground whose exact impact 
on CMB will depend.
The other possibility is that of a direct impact on CMB missions of an un-removed component
which is relevant at CMB frequencies, such as the KBOE. 
In this case the exact effect will depend on the level of contamination
compared to the level of the CMB fluctuations, on the spatial pattern, and on its combination with the CMB.

A simple argument allows us to estimate the effect of un-removed ZLE or KBOE on the CMB APS at different multipoles.
Let us to consider a sky with just ZLE or KBOE and an observatory scanning the sky to form maps.
Due to unavoidable geometrical constrains, the observed sky regions 
%of sky observed by the observatory 
are more or less tightly
correlated to specific positions in the Solar System, but it is not possible to grant 
a one--to--one correlation, so 
that when timelines are co-added to form maps the result will be a distribution 
with small discontinuities, and in general different scanning
strategies will create slightly different maps\cite{marisetal2006}\,.
However, if observations from many scans, spanning several years, 
are co-added, then a smooth map 
with a very strong planar symmetry about the ecliptic will be obtained. 
Denoting with $\almSS$ the coefficients of the spherical harmonics (SH) expansion
for such map, the planar symmetry assures that in a reference frame defined by this plane
$\almSS\equiv0$ for any odd $\ell$ or for any $m\ne0$, so that
$\ClSS={a_{\ell,0}^{\mathrm{SS}}}^2/(2\ell+1)$ for even $\ell$ and zeros for odd $\ell$.
The combined map with CMB will be unchanged for odd $\ell$ while will 
have in the same reference frame
%$\Cl=\ClCMB + \ClZLE + \sqrt{\ClZLE/(2\ell+1)} {a_{\ell,0}^{\mathrm{CMB}}}$% \cos \varphi_\ell$
$\Cl=\ClCMB + \ClSS + {a_{\ell,0}^{\mathrm{SS}}} {a_{\ell,0}^{\mathrm{CMB}}}/(2\ell+1)$.
So, depending on the signs of the ZLE or KBOE and CMB components, 
$\Cl$ can be smaller or larger than 
$\ClCMB$.  
% An also the relative alignements of peaks of multipoles will be affected.
Therefore, it can not be in principle excluded that part of the 
anomalies seen at low multipoles can be ascribed to some unknown and un-removed 
component of the ZLE or KBOE\cite{marisetal2011} and/or to interplay between this foreground and un-removed dipole-like systematic effects\cite{hansenetal2012}\,,
 especially at low $\ell$ where a Solar System large scale diffuse emission should have the maximum power.

%%%%%%%%%
% Galactic emissions

%\include{Planck}
\def\deg{\ifmmode^\circ\else$^\circ$\fi}
\def\Planck{\textit{Planck}}
\newcommand{\WMAP}{\textit{WMAP\/}}

\subsection{Galactic emissions}

%The \Planck\ mission is giving the most sensitive all-sky view of the CMB. 
%In order to obtain the true CMB distribution, the Galactic and extragalactic foregrounds must first be understood and corrected for. These foregrounds are of interest in their own right.  
The wide frequency coverage of \Planck\ when taken with relevant ancillary data provides a unique opportunity to characterize all the relevant Galactic foreground components. Of particular interest is the recently identified anomalous microwave emission (AME) due to spinning dust which has an important contribution at the lower \Planck\ frequencies.  Inclusion of this component has a domino effect on the spectrum of the other 
%three 
components, particularly at frequency $\nu \lsim 100$GHz, where synchrotron and free-free emissions are particularly important. 
%The focus of the present paper is on the plane of the inner Galaxy at lower \Planck\ frequencies ($<100$\,GHz). 
The emission from thermal (vibrational) dust dominates at $\nu \gsim 70$GHz, but, although weak, it must be considered also at lower frequencies. 
It should be remembered that the minimum of the Galactic foreground to the CMB is in the range 60--100\,GHz where each of these four components can have a small but significant contribution.     

\subsubsection{Synchrotron emission}

Synchrotron emission originates in relativistic cosmic ray electrons spiraling in the Galactic magnetic field.  The relativistic electrons are produced in the shocks associated with supernova explosions.  The spectrum of the synchrotron radio emission is related to the energy spectrum of the relativistic electrons. Up to several GHz the brightness temperature spectral index is $\sim -2.7$\,\cite{Broadbent:1989}\,; above this frequency it steepens to $-3.0$ or more at the lower \Planck\ frequencies\cite{gold2010}\,.  Another characteristic of synchrotron emission is its linear polarization which is orthogonal to the magnetic field direction;  this may be as high as 70\,\% for an aligned field with a brightness spectrum of $-3.0$.  In the more tangled field environment on the Galactic plane the observed values are in the range 10--50\,\%.

The current study of the emission of the plane in the inner Galaxy identified the synchrotron component by using component separation techniques. 
%(see below).  
The low frequency data (0.4 to 2.3\,GHz) revealed a narrow component in Galactic latitude with a FWHM of 1.6\deg.  This newly identified component is also clearly identified in K and Ka band polarization data from \WMAP; it has a brightness spectral index 
of $-3.2$\,\cite{gold2010}\,.  This narrow distribution is the sum of the supernova remnants (SNRs) over the last $10^5$--$10^6$\,years (the timescale of the SNRs before they expand into the broader latitude distribution.  A similar latitude width is found for the normal ($\sim$1 second  period) pulsars; their ages are also $\sim10^5$--$10^6$ years.  Over this timescale both the SNR shells and the pulsar proper motions will have taken them to a FWHP of 1.6\deg, double the width of the nascent OB star distribution (0.9\deg).

\subsubsection{Free-free emission}
The free-free emission in the inner Galaxy arises from the ionized (electron) gas component produced principally by the hot O and B stars which are confined to a narrow latitude width of 0.9\deg FWHM. Cooler stars also contribute to the interstellar radiation field (ISRF) which is more diffuse.  At intermediate and high latitudes the free-free emission is measured by the H$\alpha$ spectral line.  Even here a correction is needed to account for the absorption of H$\alpha$ by dust.  On the Galactic plane the dust obscuration is so great that the H$\alpha$ emission line cannot be used.  Here the radio recombination lines (RRLs) save the day.  No dust absorption correction is required.  However an electron temperature is needed to determine the emission measure (EM = $n_e^2L$) in order to derive the corresponding continuum temperature at any frequency\cite{Dickinson2003}\,.  The brightness temperature spectral index is well determined at \Planck\ frequencies; it is $\sim-2.13$ at 30\,GHz.  The electron temperature of the diffuse ionized gas appears to be similar to the average for the compact HII regions\cite{Alves2010,Alves2012}\,.  

The FWHM of the free-free (1.1\deg) is intermediate between that of OB stars (0.9\deg) and the neutral hydrogen (1.8\deg).  This is not unexpected since the gas (HI, H2 and dust) density is greatest on the plane and also because the ionized emission is proportional to $n_e^2L$. The free-free, along with the AME,  dominates the emission on the plane in the inner Galaxy.  

\subsubsection{Anomalous Microwave Emission (AME)}
AME is the recently identified emission component which is well-correlated with far-infrared (FIR) dust emission. It is produced by rapidly spinning small dust grains having an electric dipole moment\cite{Draine:1998a}\,.  Typical masses are $\sim$50 atoms which in a dust cloud produce a spectrum which peaks in the range 15--50\,GHz depending on the environment and radiation field.  \Planck\ 
%with its wide frequency coverage 
has for the first time been able to define the shape of the spectrum on the high frequency side of the emission peak in a number of dust/molecular/HII regions, as shown in [\refcite{planck2011-7.2}]. This work has provided a rich source of data to explore the emission mechanism in detail.

On the Galactic plane the AME spectrum can be estimated by applying component separation techniques to the strong signals measured here.  The AME is the residual emission after the free-free, the synchrotron and the thermal dust have been accounted for.  In the frequency range 20--40\,GHz AME is comparable in brightness to the free-free for the inner Galactic plane from $l = 300\deg$--$0\deg$--$60\deg$.  The latitude width of the emission at these frequencies is similar to that of the thermal dust.

\subsubsection{Thermal dust emission}
The FIR dust spectrum is due to the vibrational emission from dust grains heated by the ISRF.  The peak in the emission is at a wavelength of  $\sim$60--100\,microns.  Averaged over the intermediate latitude sky the dust temperature is $\sim$18\,K with a grey body slope in brightness of  $+1.7$\,\cite{planck2011-7.0}\,.  On the Galactic plane the dust temperature is somewhat higher at 20--24\,K.  The latitude width of the dust emission at say 100\,microns is 1.2\deg, similar to that of CO (representing H$_2$). The narrower width compared with HI is probably due to the higher dust temperature on the plane produced by the O and B stars.

\subsubsection{Emissions close to the Galactic plane}
We find a narrow latitude distribution on the Galactic plane for each of the four emission components, synchrotron, free-free, AME and thermal dust.  Recent star formation over the last $10^5$--$10^6$\,years in the dense gas regions on the plane is most likely the cause.  

Using precise full-sky observations from {\it Planck}, and applying several methods of component separation, the emission from the Galactic  "haze" at microwave wavelengths has been 
identified and characterized\cite{hazeplanck}\,. The haze is a distinct component of diffuse Galactic emission, roughly centered on the Galactic centre, and extends to $|b| \sim 35^{\circ}$ in Galactic latitude 
and $|l| \sim15^{\circ}$ in longitude. By combining WMAP and {\it Planck} data, [\refcite{hazeplanck}] were able to determine the spectrum of this emission to high accuracy, unhindered by the large systematic biases present in previous analyses. The derived spectrum is consistent with power-law emission with a spectral index of $-2.55 \pm 0.05$, thus excluding free-free emission as the source and instead favoring 
hard-spectrum synchrotron radiation from an electron population with a spectrum (number density per energy) $dN/dE \sim E^{-2.1}$. At Galactic latitudes $|b|<30^{\circ}$, the microwave haze morphology is consistent with that of the Fermi gamma-ray "haze" or "bubbles" (see also [\refcite{carrettietal}]), indicating that we have a multi-wavelength view of a distinct component of our Galaxy. Given both the very hard spectrum and the extended nature of the emission, it is highly unlikely that the haze electrons result from supernova shocks in the Galactic disk. Instead, a new mechanism for cosmic-ray acceleration in the centre of our Galaxy is implied. 

\smallskip 

The wide frequency coverage of \Planck, which includes polarization, allows the spectrum of each component to be determined unambiguously.  Polarization data from \Planck\ are awaited with considerable interest.

%%%%%%
% Extragalactic sources

\section{Extragalactic radio and far--IR sources at mm/sub-mm wavelengths}
\label{extra}

The {\it Planck} Early Release Compact Source Catalogue (ERCSC)\cite{Planck_Paper7} -- the {\it first complete
full-sky catalogue} of bright sub-millimeter extragalactic compact sources -- provides positions and flux densities
of hundreds of ``radio'' sources (intermediate to high--redshift Active Galactic Nuclei (AGN)) and of thousands of
``far-IR'' sources (low--redshift dusty galaxies) detected in each of the nine {\it Planck} frequency maps during
the first 1.6 {\it Planck} full-sky surveys. As shown in [\refcite{Planck_Paper7}], their Table 1, the full-sky
surveys of the {\it Planck} satellite are -- and will be, for years to come -- unique in the millimeter, at
$\lambda \leq 3$ mm, and sub-millimeter domains. Thanks to this huge amount of new data it is thus possible to
investigate the SEDs of extragalactic point sources in a spectral domain very
poorly explored before and, at the same time, their cosmological evolution, at least for some relevant source
populations.

\subsection{Radio sources: ``blazars''}

The most recent estimates on source number counts of extragalactic radio (synchrotron) sources up to $\sim50-70$
\,GHz, and the optical identifications of the corresponding point sources (see e.g. [\refcite{Massardi08}]), show
that these counts are dominated by radio sources whose average spectral index is ``flat'', i.e., $\alpha\simeq
0.0$ (with the usual convention $S_\nu\propto\nu^\alpha$). This result confirms that the underlying source
population is essentially made of Flat Spectrum Radio Quasars (FSRQ) and BL Lac objects, collectively called
``blazars''\footnote{Blazars are jet-dominated extragalactic objects characterized by a strongly variable and
polarized emission of the non-thermal radiation, from low radio energies up to high energy gamma rays; see e.g.
[\refcite{UrryPadovani95}].}, with minor contributions coming from other source populations\cite{Toffolatti98,deZotti05}\,. 
At frequencies $> 100$\,GHz, however, there is now new information for sources with
flux densities below about $1\,$Jy, coming from the South Pole Telescope (SPT) collaboration\cite{Vieira10}\,, with
surveys over 87 deg$^2$ at 150 and 220\,GHz, and from the Atacama Cosmology Telescope (ACT) survey over 455
deg$^2$ at 148\,GHz\,\cite{Marriage11}\,.

To study the spectral properties of the extragalactic radio sources in the {\it Planck} ERCSC\cite{Planck_Paper13} used a reference 30\,GHz sample above an estimated completeness limit $S_{lim}\simeq
1.0\,$Jy. In this sample, the 30--143\,GHz median spectral index is in very good agreement with the one found by
Marriage et al.\cite{Marriage11} for their bright ($S_\nu>50$ mJy) 148\,GHz-selected sample with complete
cross-identifications from the Australia Telescope 20\,GHz survey, i.e $\alpha_{20}^{148}=-0.39\pm 0.04$. In the
whole, the results of [\refcite{Planck_Paper13}] show that in their sample selected at 30 GHz a moderate steepening
of the spectral indices of the radio sources at high radio frequencies, i.e. $\geq 70-100$ GHz, is clearly
apparent. It has also been shown by [\refcite{Planck_Paper13}] that differential number counts at 30, 44, and
70\,GHz are in good agreement with those derived from \textit{WMAP} data\cite{wright09} at nearby frequencies.
The model proposed by de Zotti et al.\cite{deZotti05} in 2005 is consistent with the present counts at frequencies
up to 70\,GHz, but over-predicts the counts at higher frequencies by a factor of about 2.0 at 143\,GHz and about
2.6 at 217\,GHz. As reminded before, the analysis of the spectral index distribution over different frequency
intervals, within the uniquely broad range covered by {\it Planck} in the mm and sub-mm domain, has highlighted an
average {\it steepening} of source spectra above about 70\,GHz. This steepening accounts for the discrepancy
between the model predictions of de Zotti et al.\cite{deZotti05} and the observed differential number counts at
HFI frequencies.

%Moreover, in \refcite{Planck_Paper15} a detailed discussion on the modelling of the spectra of blazars is also
%presented. In this paper, spectral energy distributions (SEDs) and radio continuum spectra are presented for a
%northern sample of 104 extragalactic radio sources, based on the Planck Early Release Compact Source Catalogue
%(ERCSC) and simultaneous multifrequency data. The nine Planck frequencies, from 30 to 857 GHz, are complemented by
%a set of quasi--simultaneous observations ranging from radio to gamma-rays. SED modelling methods are discussed,
%with an emphasis on proper, physical modelling of the synchrotron bump using multiple components, and a thorough
%discussion on the original accelerated electron energy spectrum in blazar jets is presented. The main conclusion
%is that, al least for a fraction of the observed mm/sub-mm blazar spectra, the energy spectrum could be much
%harder than commonly thought, with a power-law index $\sim 1.5$ and the implications of this hard value are
%discussed for the acceleration mechanisms effective in blazar shocks.

Recently, a successful explanation of the change detected in the spectral behavior of extragalactic radio sources
(ERS) at frequencies above 70-80\,GHz has been proposed by Tucci et al.\cite{Tucci11}\,. By applying the
K\"onigl\cite{Konigl81} model for the emission in the inner jets of blazars, [\refcite{Tucci11}] makes a first attempt at
constraining the most relevant physical parameters that characterize the emission of blazar sources by using the
number counts and the spectral properties of extragalactic radio sources estimated from high--frequency radio
surveys\footnote{The main goal of [\refcite{Tucci11}] was to present physically grounded models to extrapolate the
number counts of ERS, observationally determined over very large flux density intervals at cm wavelengths down to
mm wavelengths, where experiments aimed at accurately measuring CMB anisotropies are carried out.}. As noted
before, a relevant steepening in blazar spectra with emerging spectral indices in the interval between $-0.5$ and
$-1.2$, is commonly observed at mm/sub-mm wavelengths. Tucci et al.\cite{Tucci11} interpreted this spectral
behavior as caused, at least partially, by the transition from the optically--thick to the optically--thin regime
in the observed synchrotron emission of AGN jets\cite{Marscher96}\,, giving rise to a ``break'' frequency, $\nu_M$,
typically in the range between 50-2000 GHz, at which the synchrotron spectrum of jets bends down\footnote{Based on
published models, Tucci et al.\cite{Tucci11} estimated the value of the frequency $\nu_M$ (and of the
corresponding radius $r_M$) at which the break occurs on the basis of the ERS flux densities measured at 5\,GHz
and of the most typical values for the relevant physical parameters of AGN jets.}. On the whole, the results of
[\refcite{Tucci11}] imply that the parameter $r_M$ should be of parsec--scales, at least for FSRQs, in agreement
with theoretical predictions\cite{Marscher85}\,, whereas values of $r_M\ll 1\,$pc should be only typical of BL\,Lac
objects or of rare, and compact, quasar sources.

\subsection{Far--IR sources: local dusty galaxies}

The analysis done by [\refcite{Planck_Paper16}] presented the first results on the properties of nearby galaxies
using ERCSC data. From reliable associations between {\it Planck} and IRAS, they selected a subset of 468 for
SED studies, namely those with strong detections in the three highest frequency
{\it Planck} bands and no evidence of cirrus contamination. This selection has thus provided a first {\it Planck} sample
of local, i.e. at redshift $< 0.1$, dusty galaxies \footnote{This sample is very important for determining their
emission properties and, in particular, the presence of different dust components contributing to their sub-mm
SEDs.}. The analysis of SEDs of these local galaxies\cite{Planck_Paper16} has confirmed the presence of cold dust
in local giant and, largely, in dwarf galaxies\footnote{The SEDs are fitted using parametric dust models to
determine the range of dust temperatures and emissivities. They found evidence for colder dust than has previously
been found in external galaxies, with temperatures $T < 20$ K. Such cold temperatures are found by using both the
standard single temperature dust model with variable emissivity $\beta$, or a two dust temperature model with
$\beta$ fixed at 2.}. In [\refcite{Planck_Paper16}] it is also found that some local galaxies are both luminous and
cool, with properties similar to those of the distant SMGs uncovered in deep sub-mm surveys. The main conclusion of
[\refcite{Planck_Paper16}] is that cold ($T < 20$ K) dust is thus a significant and largely unexplored component of
many nearby galaxies and that there is a new population of cool sub-mm galaxies, showing the presence of even
cooler dust grains, with estimated temperatures of $\sim$10-13 K.

%\begin{figure}[t]
%\begin{center}
%\psfig{file=diff-number-counts-VII.eps,width=6in}
%\end{center}
%\vskip -0.4cm \caption{Fig. 9 from \cite{Planck_IntPap7}. {\it Planck}
%   differential number counts, normalized to the
%    Euclidean value (i.e.\ $S^{2.5} dN/dS$), compared with models and
%    other data sets. {\it Planck} counts: total (black filled circles);
%    dusty (red circles); synchrotron (blue circles). Four models are
%    also plotted: \cite{deZotti05}, dealing only with synchrotron sources -- solid line;
%    \cite{Tucci11} dealing only with synchrotron sources -- dots;
%    \cite{Bethermin11} dealing only with dusty sources -- long dashes;
%    \cite{Serjeant05} dealing only with local dusty sources -- short dashes.
%    Other data sets:
%    \Planck\ early counts for 30\,GHz-selected radio galaxies
%    \cite{Planck_Paper13} at 100, 143 and 217\,GHz (open diamonds);
%    {\it Herschel} ATLAS and HerMES counts at 350 and 500\micron\ from
%    \cite{Oliver10} and \cite{Clements10}; BLAST at the same two
%    wavelengths, from \cite{Bethermin10}, all shown as
%    triangles. Left vertical axes are in units of
%    Jy$^{1.5}$\,sr$^{-1}$, and the right vertical axis in
%    Jy$^{1.5}$.deg$^{-2}$. Credit: Planck Collaboration, A\&A, submitted (ms AA/2012/20053), 2012, reproduced with
%permission \copyright\ ESO.} \label{F12}
%\end{figure}

Very recently, using selected samples from the first {\it Planck} 1.6 full-sky surveys, i.e. the {\it Planck}
ERCSC, [\refcite{Planck_IntPap7}] derived number counts of extragalactic point sources from 100 to 857\,GHz
(3\,mm to 350 $\mu$m). More specifically, for the first time, number counts have been provided of synchrotron
dominated sources (blazars) at high {\it Planck} frequencies (353 to 857\,GHz) and of dusty galaxies at lower
frequencies (217 and 353\,GHz). {\it Planck} number counts are found to be in the Euclidean regime in this
frequency range, since the ERCSC comprises only bright sources ($S> 0.3$ Jy). The estimated number counts appear
generally in agreement with other data sets, when available (see [\refcite{Planck_IntPap7}] for more details).

These new estimates of number counts of synchrotron and of dust--dominated extragalactic sources allowed new
constraints to be placed on cosmological evolution models which extend their predictions to bright flux densities,
i.e. $S> 1$ Jy. A very relevant result is that the most successful model of Tucci et al.\cite{Tucci11} is
performing particularly well at reproducing the number counts of synchrotron--dominated sources up to 545\,GHz. On
the contrary, [\refcite{Planck_IntPap7}] highlights the failure of many models for number count predictions of dusty
galaxies to reproduce all the high-frequency counts. The likely origin of these discrepancies is an inaccurate
description of the galaxy SEDs used at low redshift in these models. Indeed a cold dust component, detected by
[\refcite{Planck_Paper16}]\,, is rarely included in the models of galaxy SEDs at low redshift. On the whole, these
results already obtained by the exploitation of the {\it Planck} ERCSC data are providing valuable information
about the ubiquity of cold dust in the local Universe, at least in statistical terms, and are guiding to a better
understanding of the cosmological evolution of extragalactic point sources at mm/sub-mm wavelengths.

%%%%%%
% DePaolis

\begin{figure}[pb]
\centerline{\psfig{file=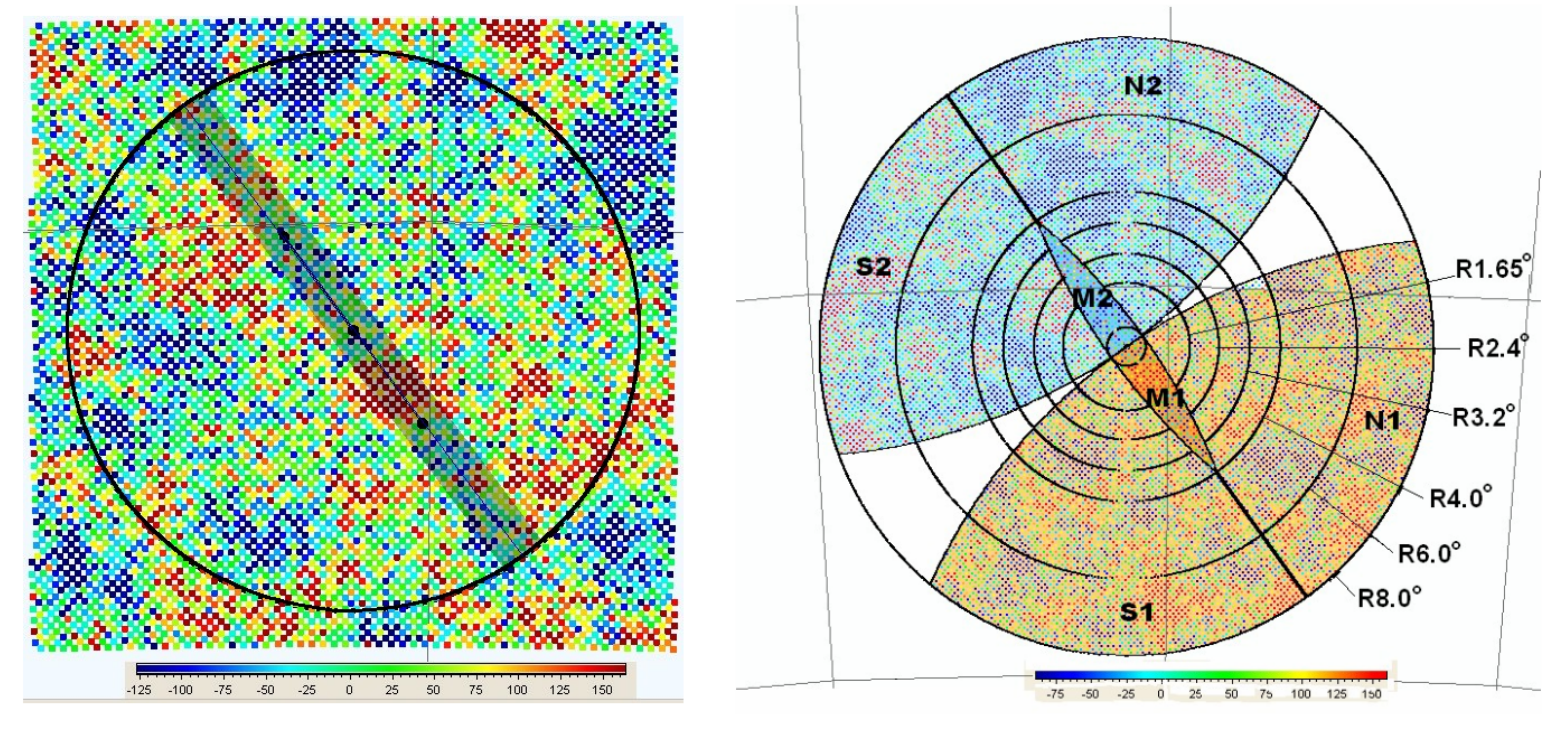,width=10.cm}}
\vspace*{8pt}
\caption{The detailed geometry (up to 8\deg)
used in the analysis is shown. The different pixel colors indicate the difference of the CMB temperature with
respect to the average temperature set to zero. Red means positive
excess and goes up to a maximum of 150$\mu$K while blue means
lower temperature and goes up to $-125$$\mu$K. The left image shows
the real WMAP W band map, while the right image shows the geometry used
in the analysis and the average temperature in the two sides of the M31
disk and halo in false colors. It shows in a single glance that one side
of both the M31 disk and halo is hotter with respect to the other side.\label{f1}}
\end{figure}

\subsection{Nearby galaxies: the case of M31}

%CMB data, such as those obtained by  the WMAP and {\it Planck} satellites, are generally used to constrain the cosmological parameters. 
%However, these data can also be used to get local information about the CMB sky, e.g. detecting  point like astrophysical sources (see e.g. [\refcite{gurzadyan}]  and references therein). 
As discussed above, WMAP and {\it Planck} data can be used to get information about point like astrophysical sources (see e.g. [\refcite{gurzadyan}]  and references therein), including 
nearby galaxies.
Recently, the  7-year WMAP data have been used to trace the disk and the halo of 
the M31 galaxy\cite{depaolis2011}\,. Unexpectedly, an asymmetry in the mean microwave temperature towards both the M31 disk and halo, along the direction of the M31 rotation, has been found.
The maximum temperature contrast (see Fig.~\ref{f1}, reprinted from [\refcite{depaolis2011}]) is  about 130 $\mu$K/pixel (or about 200 
$\mu$K/pixel if the M31 Bulge is excluded). This temperature asymmetry, similar in the three WMAP bands W, V and Q, is very likely induced by the Doppler shift effect due to the M31 disk rotation speed. A similar effect is clearly visible also towards the M31 halo up to about 120 kpc from the M31 center with a peak value of about 40  $\mu$K/pixel.

The robustness of this result has been tested, both for the M31 disk and halo, by considering 500 randomly distributed control fields in the three WMAP bands and also by simulating 500 sky maps 
(see [\refcite{depaolisMG13}] for more details). CMB maps are simulated by assuming $\Delta T(\hat{n}) = \Delta T_{CMB}(\hat{n})\otimes B(\hat{n}) + N(\hat{n})$, where $\Delta T_{CMB} $ is a realization of the Gaussian CMB field, $N(\hat{n})$ is the pixel noise and $B(\hat{n})$ is the proper beam of the experiment. Using the synfast routine of HEALPix\cite{gorski} with the best-fit power spectrum constrained with BAO and $H_0$, as given by the WMAP Collaboration, 500 realizations of the CMB sky were made. The maps have been then convolved with the WMAP beams for W, V, and Q bands, respectively,  taking into account the convolution with the beam function of the experiment and randomly extracting the noise value from a normal distribution with 
$\sigma=\sigma_0/\sqrt{N_{obs}}$. The statistical analysis shows that there is a probability below about $1\% $ that the temperature asymmetries both in the M31 disk and halo are due to random fluctuations of the CMB signal. 

The degree to which galactic halos rotate with respect to the disks is a difficult task to be investigated.  In this respect, the methodology of using CMB data to probe both the disk and the halo of M31, even if with the limitation of the presently available data, may suggest a novel way of approaching this problem especially in view of the high accuracy CMB measurements with the {\it Planck} satellite.

%%%%%%
% CIB re-adapted from Toffolatti et al. 2012

\subsection{Cosmic Infrared Background anisotropies}

The Cosmic Infrared Background (CIB) is the relic emission, at wavelengths larger than a few microns, of the
formation and evolution of the galaxies of all types, including Active Galactic Nuclei (AGN) and star-forming
systems \cite{Puget96,buriganaetal97,buriganapopa98,Hauser01,Dole06}\footnote{An important goal of studies about galaxy formation has thus been
the characterization of the statistical behavior of galaxies responsible for the CIB - such as the number counts,
redshift distribution, mean SED, luminosity function, clustering -- and their physical properties, such as the
roles of star-forming vs. accreting systems, the density of star formation, and the number density of very hot
stars.}. The CIB accounts for roughly half of the total energy in the optical/infrared Extragalactic Background
Light (EBL)\cite{Hauser01}\,, although with some uncertainty, and its SED peaks near 150 $\mu$m. Since local
galaxies give rise to an integrated infrared output that amounts to only about a third of the optical one\cite{Soifer91}\,, 
there must have been a strong evolution of galaxy properties towards enhanced far--IR output in
the past. Therefore, the CIB, made up by high density, faint and distant galaxies\footnote{The CIB records much of
the radiant energy released by processes of structure formation occurred since the decoupling of matter and
radiation, four hundred thousand years after the Big Bang, when the CMB was produced.} is barely resolved into its
constituents. Indeed, less than 10\% of the CIB is resolved by Spitzer at 160 $\mu$m\,\cite{Bethermin10}\,, 
 $\sim 10$\% by {\sl Herschel} at 350 $\mu$m\,\cite{Oliver10}\, and $\sim 16$\% by
   the SCUBA-2 Cosmology Legacy Survey (S2CLS) at 450 $\mu$m\,\cite{Geachetal2012}\,.
%Thus, in the absence of foreground (Galactic dust) and CMB
%emissions, and when the instrument noise is subdominant, maps of the diffuse emission at the angular resolution
%probed by the current surveys reveal a web of structures, characteristic of CIB anisotropies. 
With the advent of large area far-IR to millimeter surveys ({\sl Herschel}, {\it Planck}, SPT, and ACT), CIB anisotropies thus
constitute a new tool for structure formation and evolution studies.

%CIB anisotropies are expected to trace large-scale structures and probe the clustering properties of galaxies,
%which in turn are linked to those of their hosting dark matter halos. 
Because the clustering of dark matter is reasonably well understood,
observations of anisotropies in the CIB constrain the relationship between dusty, star-forming
galaxies at high redshift, i.e. $z > 2$, and the underlying dark matter distribution.
%The connection between a population of galaxies and dark matter halos can be described by its Halo Occupation
%Distribution (HOD) (\cite{Peacock00,Berlind02}), which specifies the probability distribution of the number of
%objects with a given property (e.g., luminosity, stellar mass, or star--formation rate) within a dark matter halo
%of a given mass and their radial distribution within the halo\footnote{The HOD and the halo model provide a
%powerful theoretical framework for describing the connection between galaxies and dark matter halos. Once
%decisions are made about which properties of the halos and their environment the HOD depends upon, what the
%moments of the HOD are and what the radial profile of objects within halos is, the halo model can be used to
%predict any clustering-related observable.}. In particular, the halo model predicts that the bias, describing the
%clustering of galaxies in relation to the dark matter, becomes scale-independent at large scales. This assumption
%of a scale-independent bias is often made in modelling the CIB.
The APS of CIB anisotropies has two contributions: a white-noise component caused by shot noise
and an additional component caused by spatial correlations between the sources of the CIB. Correlated CIB
anisotropies have already been measured by many space--borne as well as ground--based experiments (see
[\refcite{Planck_Paper18}] for more details). 
%Depending on the frequency, the angular resolution and size of the
%survey, these measurements can probe two different clustering regimes. 
On small angular scales ($\ell\geq  2000$),
they measure the clustering within a single dark matter halo and, accordingly, the physics governing how dusty,
star--forming galaxies form within a halo. On larger angular scales, i.e. $200\leq\ell\leq 2000$, CIB anisotropies
measure clustering between galaxies in different dark matter halos. These measurements primarily constrain the
large-scale, linear bias, b, of dusty galaxies, which is usually assumed to be scale-independent over the relevant
range.

%Given their limited dynamic range in scale, current measurements are equally consistent with an HOD model, a
%power-law correlation function or a scale-independent, linear bias.

Thanks to the exceptional quality of the {\it Planck} data, [\refcite{Planck_Paper18}] were able to measure the
clustering of dusty, star-forming galaxies at 217, 353, 545, and 857 GHz with unprecedented precision. 
%The CIB
%maps were cleaned using templates: HI for Galactic cirrus; and the Planck 143 GHz maps for CMB. Having HI data is
%necessary to cleanly separate CIB and cirrus fluctuations. 
After careful cleaning, 
based on suitable templates and {\it Planck} maps,
they obtained CIB anisotropy
maps that reveal structures produced by the cumulative emission of high-redshift, dusty, star--forming galaxies.
%The maps are highly correlated at high {\it Planck} frequencies whereas de-correlate at lower {\it Planck} HFI
%frequencies. 
The power spectra of the latter maps 
%and their associated errors using a
%dedicated pipeline and ended up with measurements of the APS of the CIB anisotropy, $C_{\ell}$, at 217, 353, 545,
%and 857 GHz, 
were then computed with high signal-to-noise ratio over the range $200 < l < 2000$ by [\refcite{Planck_Paper18}].
These measurements compare very well
with previous measurements at higher $\ell$\footnote{The SED of CIB anisotropies is not different from the CIB
mean SED, even at 217 GHz. This is expected from the model of [\refcite{Bethermin11}] and reflects the fact that the
CIB intensity and anisotropies are produced by the same population of sources.}. Moreover, from {\it Planck} data 
alone [\refcite{Planck_Paper18}] could exclude a model where galaxies trace the
(linear theory) matter power spectrum with a scale-independent bias: that model requires an {\it unrealistic} high
level of shot noise to match the small-scale power they observed. Consequently, an alternative model that couples
the dusty galaxy, parametric evolution model of [\refcite{Bethermin11}] with a halo model approach has been developed
(see, again, [\refcite{Planck_Paper18}] for more details). Characterized by only two parameters, this model provides an
excellent fit to the measured CIB anisotropy APS for each frequency treated independently. 
%In the next future, modeling and interpretation of the CIB anisotropy will be aided by the use of cross-power spectra
%between bands, and by the combination of the Planck and Herschel data at 857 and 545/600 GHz and Planck and
%SPT/ACT data at 220 GHz.

\def\xmm{{\it XMM-Newton}}
\def\chandra{{\it Chandra}}
\def\planck{{\it Planck}}
\def\Planck{{\it Planck}}
\def\rosat{{\it ROSAT}}
\def\rass{{\rm RASS}}

%%%%%%
% Clusters 1

\section{Clusters of galaxies and their cosmological implications}
\label{clusters}

The observation of clusters of galaxies through the Sunyaev-Zel'dovich (SZ) effect, the inverse Compton scattering of cosmic microwave photon by hot intra-cluster electrons\cite{sunyaev72}\,, have proven to be an efficient way to search for new clusters\cite{Marriage11,car11,planck2011-5.1b}\,. 
%clusters\cite{mar11,car11,planck2011-5.1b}\,. 

The {\it Planck} satellite has been observing clusters of galaxies via the measurement of the SZ effect over the whole sky since August 2009. 
Although, its spatial resolution is moderate with respect to ground based SZ surveys (see e.g. [\refcite{Marriage11,car11}]), 
%[\refcite{mar11,car11}]), 
it possesses a unique nine-band coverage from 30 to 857\,GHz and, most crucially, it covers an exceptionally large survey volume. Indeed {\it Planck} is the first all-sky survey capable of blind cluster detections since the \rosat\  All-Sky Survey (\rass, in the X-ray domain). Early {\it Planck}  results on galaxy clusters were recently published in [\refcite{planck2011-5.1b,planck2011-5.1a,planck2011-5.2a,planck2011-5.2b,planck2011-5.2c,planck2011-5.1c}]. These results include the publication of the high signal--to--noise ratio (${\rm S/N} > 6$) Early SZ (ESZ) cluster sample\cite{planck2011-5.1a}\,. 

\subsection{{\it Planck} SZ clusters}

Using this specific SZ signature,  {\it Planck} was designed to be able to
detect numerous clusters\cite{agh97}\,. Unfortunately, not all are
showing up as Abell 2319. The signal is indeed quite weak and is
contaminated by foregrounds (our Galaxy, and nearby radio/IR galaxies)
and backgrounds (CMB and CIB). As described later, the published
{\it Planck} clusters have a signal--to--noise ratio (S/N) greater than
6. This means that the S/N per frequency is of the order of 1. This
has lead us to develop a specific approach for detecting, validating
and confirming clusters.

We use a multi-matched filter (MMF) method\cite{mel06} to detect
the clusters. It is taking advantage of the spectral signature (SZ
signature without relativistic effects) {\it and } the spatial
signature (universal spherical profile from X-ray REXCESS observations\cite{arn10}) 
of the clusters detected by {\it Planck}. As optimal as the
method can be, a process of validation is still necessary to remove
false detections. This is done in two steps. First a cross-check with
internal {\it Planck} catalogues (cold cores, solar system objects, bad
pixels) is performed, then cross-checks with existing external
catalogues and data (SDSS, RASS) are performed to classify the known
clusters and the new candidate clusters.  Finally, follow-up
observations has been done in optical, SZ and mainly in X-ray with
XMM-Newton, to confirm our candidate clusters.

\subsubsection{{\it Planck} Early SZ cluster sample}

\begin{figure}[t]
 \centering
 \includegraphics[width=0.6\textwidth,clip]{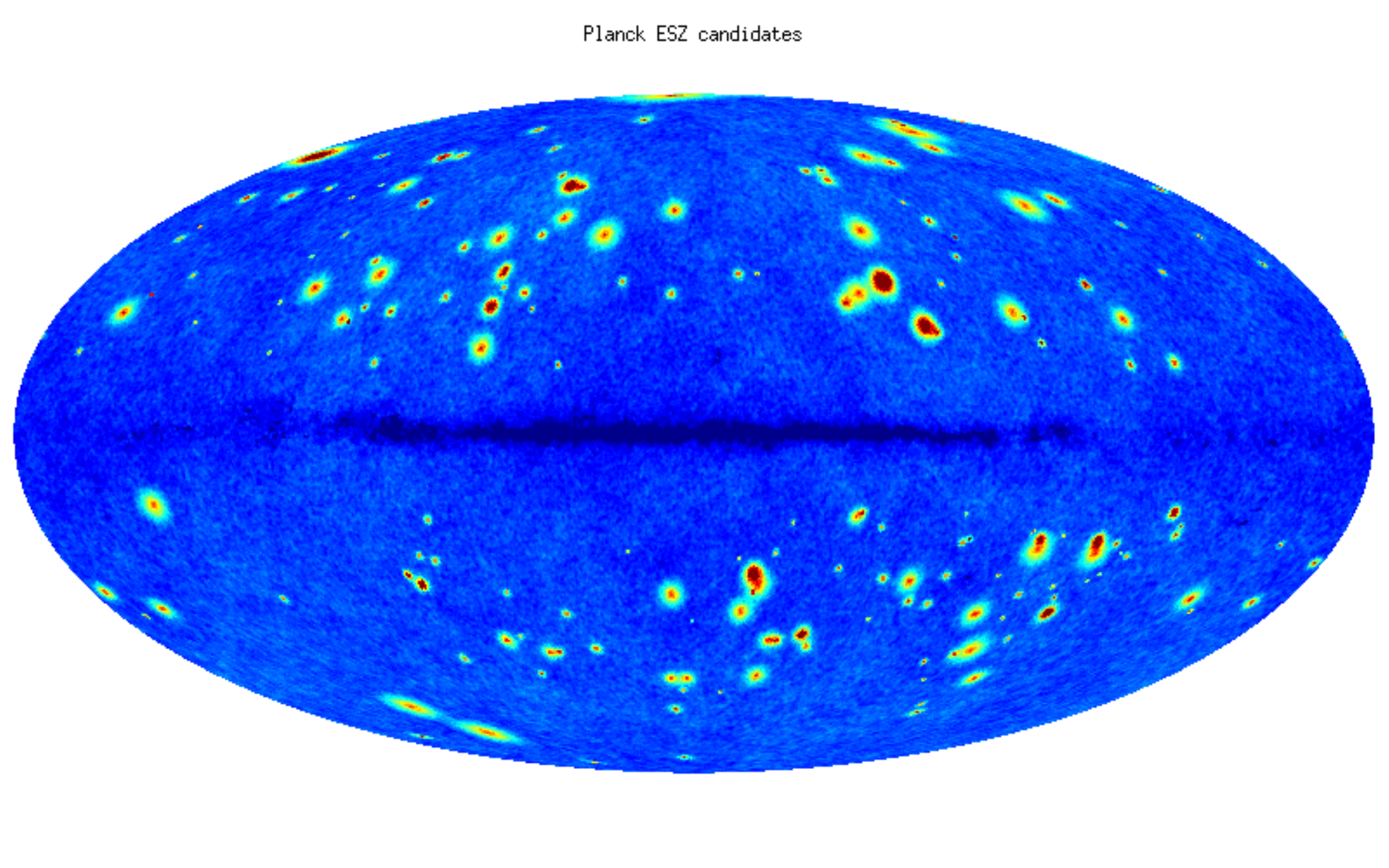}
% {douspis_fig4}
%% Note the ABSENCE of the extension .pdf , .eps or .ps  !
  \caption{Distribution on the sky of the {\it Planck} ESZ clusters (the signal has been amplified to be seen).}
  \label{author1:fig4}
\end{figure}

These detection, validation, and confirmation steps have lead to the
production of the {\it Planck} Early SZ Cluster sample (ESZ). It contains
199 clusters, 10 of which, confirmed by XMM-Newton validation program\cite{planck2011-5.1b, planck2012-I, planck2012-IV}\,,
have a S/N $<$6. The 189 clusters with S/N greater than 6 are divided in 169 known
clusters (in X-ray, optical or SZ) and 20 new {\it Planck} clusters. At the
time of the release only 11 were confirmed by XMM-Newton. Since then,
6 more have been confirmed by SPT and AMI\cite{sptesz, amiesz}\,.
The sample is available\footnote{rssd.esa.int/Planck} as part of the {\it Planck} Early Release Compact
Source Catalogue (ERCSC)\cite{Planck_Paper7}\,.
%\cite{planck2011-1.10} 
%at rssd.esa.int/Planck. 
The distribution on the sky of these clusters is
shown in Fig.~\ref{author1:fig4} (reprinted from [\refcite{douspis_conf}]).

The ESZ clusters have relatively low redshift; 86\% of them have $z <
0.3$. Their masses span more than a decade up to $1.5 \cdot 10^{15}
M_{sol}$, and a large fraction of new {\it Planck} detected clusters are
massive ($>9 \cdot 10^{14} M_{sol}$). {\it Planck} has thus a unique capability to
detect the rarest and most massive clusters over the full sky.

\subsubsection{SZ clusters properties}

Observing galaxy clusters in SZ opens a new observational window to
understand 
%not only 
the clusters themselves 
%but also 
and the evolution of
our Universe. 
%As described earlier, 
{\it Planck} has detected new clusters,
sometimes massive. Why have they not been detected already in X-ray?
Is this a new population of clusters, or the gas (responsible for both
X-ray and SZ emissions) properties differ from what we think? As
massive objects, clusters are sensitive to cosmological initial
conditions and cosmic evolution. To use clusters for cosmological
studies we need to relate their mass to our observation (SZ effect or
Y-parameter). But is SZ effect a good proxy for the mass? How does the SZ
signal relates to the X-ray luminosity, to the richness of clusters?
The {\it Planck} ESZ clusters and {\it Planck} data are and will help in answering
these questions.

\subsubsection{New {\it Planck} clusters}

\begin{figure}[t]
 \centering
  \includegraphics[width=0.9\textwidth,clip]{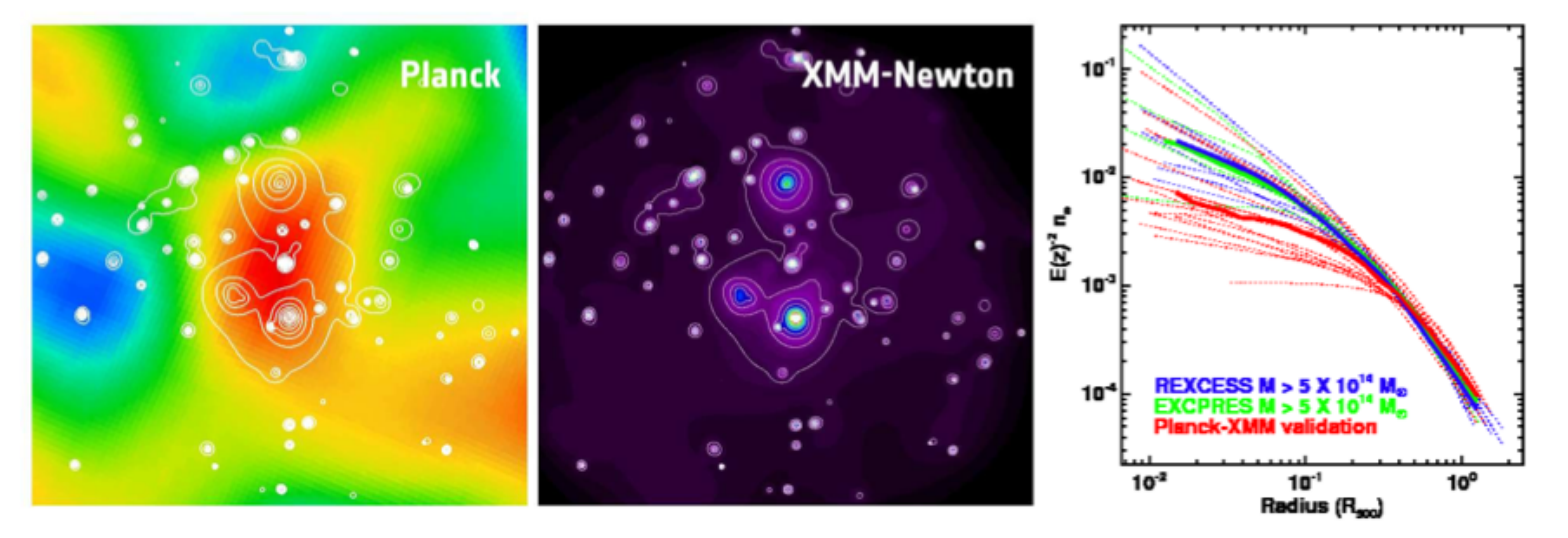}%      
% \includegraphics[width=0.6\textwidth,clip]{douspis_fig5.pdf}%      
% \includegraphics[width=0.28\textwidth,clip]{douspis_fig6.pdf}%      
%% Note the ABSENCE of the extension .pdf , .eps or .ps  !
  \caption{{\bf Left panel:} {\it Planck} SZ and XMM X-ray images of PLCKG214.6+37.0. {\bf Right panel:} electronic density profiles of {\it Planck} and REXCESS clusters.}
  \label{author1:fig5}
\end{figure}

The new {\it Planck} confirmed clusters have been compared with REXCESS
X-ray detected clusters. {\it Planck} clusters show a more complex
morphology, being sometimes really diffuse, extended, disturbed, and
also double or triple. For the same given mass, they are also
sub-luminous in X-ray compared to the REXCESS ones. Their electronic
density profiles is on average lower in the center than the REXCESS
ones (see Fig.~\ref{author1:fig5}, reprinted from [\refcite{planck2011-5.1b}]). Multi-wavelenght studies will help understand these properties. For
example, [\refcite{bag11}] have observed one the XMM confirmed {\it Planck} new
clusters and found radio arcs. Such findings, revealing shocks and/or
merger, would imply higher temperature areas, that could enhance the
SZ signal and explain why these clusters are seen in SZ and not in
X-ray. More dedicated multi-wavelenght studies are thus needed to
better understand these clusters.

%%%%%%
% Clusters 2

%Block defining journal reference macros: Ap.J. style.
%
\def\etal{et {\it al.} }
\def\araa{{\it Ann.\ Rev.\ Astron.\ Ap.}}
\def\aplet{{\it Ap.\ Letters}}
\def\aj{{\it Astron.\ J.}}
\def\apj{ApJ}
\def\apjl{{\it ApJ\ (Lett.)}}
\def\apjs{{\it ApJ\ Suppl.}}
\def\aas{{\it Astron.\ Astrophys.\ Suppl.}}
\def\aa{{\it A\&A}}
\def\aap{{\it A\&A}}
\def\mnras{{\it MNRAS}}
\def\nat{{\it Nature}}
\def\pasa{{\it Proc.\ Astr.\ Soc.\ Aust.}}
\def\pasp{{\it P.\ A.\ S.\ P.}}
\def\pasj{{\it PASJ}}
\def\pre{{\it Preprint}}
\def\prd{{\it Phy.\ Rev.\ D}}
\def\sovlet{{\it Sov. Astron. Lett.}}
\def\adspr{{\it Adv. Space. Res.}}
\def\expas{{\it Experimental Astron.}}
\def\ssr{{\it Space Sci. Rev.}}
\def\apss{{\it Astrophys. and Space Sci.}}
\def\inpress{in press.}
\def\souspresse{sous presse.}
\def\inprep{in preparation.}
\def\enprep{en pr\'eparation.}
\def\submit{submitted.}
\def\soumis{soumis.}
\def\aph{{\it Astro-ph}}
\def\astroph{{\it Astro-ph}}
\def\physrep{{\it Physics Reports}}
% Some other macros used in the sample text
\def\st{\scriptstyle}
\def\sst{\scriptscriptstyle}
\def\mco{\multicolumn}
\def\epp{\epsilon^{\prime}}
\def\vep{\varepsilon}
\def\ra{\rightarrow}
\def\ppg{\pi^+\pi^-\gamma}
\def\vp{{\bf p}}
\def\ko{K^0}
\def\kb{\bar{K^0}}
\def\al{\alpha}
\def\ab{\bar{\alpha}}
\def\be{\begin{equation}}
\def\ee{\end{equation}}
\def\bea{\begin{eqnarray}}
\def\eea{\end{eqnarray}}
\def\CPbar{\hbox{{\rm CP}\hskip-1.80em{/}}}
%temp replacement due to no font
%
%\def\ergs{ergs s$^{-1}$}
\def\ergscm{\rm ergs\,s^{-1}\,cm^{-2}}
\def\pho{ photons cm$^{-2}$ s$^{-1}$ }
\def\phokev{ photons cm$^{-2}$ s$^{-1}$ keV$^{-1}$}
\def\ap{$\approx$ }
\def\ep{$\rm e^\pm$ }
\def\mjyb{mJy/beam }
%Block defining astrophysical quantities
%
% SZ
\def\vp{{ v_{p}}}
\def\te{{T_{e}}}
\def\rc{{r_{c}}}
\def\yc{{y_{c}}}
\def\neu{{n_{e}}}
% Physical constants
\def\hzero{{H_{0}}}
\def\qzero{{q_{0}}}
\def\tcmb{{T_{cmb}}}
\def\sigtom{{\sigma_{T}}}
% divers
\def\cd{{C_{dust}}}
\def\td{{T_{dust}}}
\def\nd{{n_{dust}}}
\def\inu{{I_{\nu}}}
\def\fnu{{F_{\nu}}}
\def\bnu{{B_{\nu}}}
\def\mecdeux{{m_{e}c^{2}}}
\def\msol{{M$_{\odot}$}}
\def\xmm{XMM-{\it Newton} }
\newcommand{\ccor}[1]{\textcolor{black}{#1}}
\newcommand{\ccog}[1]{\textcolor{black}{#1}}
\newcommand{\ccom}[1]{\textcolor{blue}{#1}}
\newcommand{\ccop}[1]{\textcolor{pink}{#1}}
\def\xmm{{\it XMM-Newton}}
\def\chandra{{\it Chandra}}
\def\planck{{\it Planck}}
\def\Planck{{\it Planck}}
\def\rosat{{\it ROSAT}}
\def\rass{{\rm RASS}}
\def \xspec {\hbox{\tt xspec}}
\def \evigweight {\hbox{\sc evigweight}}
\def \asmooth{\hbox{\sc asmooth}}
\def \sas {\hbox{SAS}}
\def \wabs {\hbox{\sc wabs}} 
\def \epic {\hbox{\sc EPIC}} 
\def \mos {\hbox{\sc EMOS}} 
\def \pn {\hbox{\sc EPN}} 
\def \mekal {\hbox{\sc mekal}}
\newfont{\gwpfont}{cmssq8 scaled 1000}
\newcommand{\rexcess}{{\gwpfont REXCESS}}
\newcommand{\excpres}{{\gwpfont EXCPRES}}
\newcommand{\reflex}{{REFLEX}}
\newcommand{\noras}{{NORAS}}
\newcommand{\bcs}{{BCS}}
\newcommand{\ebcs}{{eBCS}}
\newcommand{\macs}{{MACS}}
%
%\renewcommand{\textfraction}{0.001}	% allow minimal text w. figs
% divers gwp
\def\Mv{M_{500}}
\def\Rv{R_{500}}
\def\Mgv{M_{\rm g,500}}
\def\Mg{M_{\rm g}}
\def\YX {Y_{\rm X}}
\def\TX {T_{\rm X}}
\def\kT {{\rm k}T}
\def\YSZ {Y_{\rm SZ}}
\def\YSZ {Y_{500}}
\def\fgv {f_{\rm g,500}}
\def\fg  {f_{\rm g}}
\def\kT {{\rm k}T}
\def\Mv {M_{\rm 500}}
\def \Rv {R_{500}}
\def\keV {\rm keV}
\def\Yv {Y_{500}}
\def\LX {L_{500,[0.1-2.4]\,\keV}}
\def\MT {$M_{500}$--$T_{\rm X}$}
\def\MYX {$M_{500}$--$Y_{\rm X}$}
\def\MMg {$M_{500}$--$M_{\rm g,500}$}
\def\MgT {$M_{\rm g,500}$--$T_{\rm X}$}
\def\MgY {$M_{\rm g,500}$--$Y_{\rm X}$}
\def\Lxz{$L_{\rm X}$--$z$}
\def\YXM {$\YX$--$\Mv$}
\def\YSZYX {$\YSZ$--$\YX$}
\def\YXYSZ{$\YX$--$\YSZ$}
\def\LXM {$\LX$--$\Mv$}
\def\fgas {$f_{\rm gas}$}
\def\toto{PLCK\,G266.6$-$27.3}
\def\msol {{\rm M_{\odot}}}
\def\lesssim{\mathrel{\hbox{\rlap{\hbox{\lower4pt\hbox{$\sim$}}}\hbox{$<$}}}}
\def\gtrsim{\mathrel{\hbox{\rlap{\hbox{\lower4pt\hbox{$\sim$}}}\hbox{$>$}}}}

\newcommand{\propsim}{\lower 3pt \hbox{$\, \buildrel \frac{{\textstyle \propto}{\textstyle \sim}}  \,$}}
\subsection{Baryons in clusters of galaxies as seen in the \Planck\ survey}

The total SZ signal is closely related to the cluster mass (see e.g. [\refcite{das04}]), and its surface brightness insensitive to distance. Therefore, SZ surveys can potentially be used to built unbiased close to mass selected cluster samples up to high redshift. These scaling relations also bear the imprint of all gravitational and non-gravitational physical processes at play in the process of structure formation and evolution.
Therefore such SZ samples of galaxy clusters will be of tremendous help for structure formation studies and to provide CMB independent cosmological constraints 
(see e.g. [\refcite{seh10,benson11,reichardt12}]. 
However, this requires a precise understanding of the statistical properties of the cluster population and furthermore a precise calibration of scaling relations  between clusters physical properties and their mass. 
In the following, we focus on the  current results on SZ scaling relations with respect to \Planck's results.  From three different approaches, we have brought tight constraints on the scaling relations between the SZ signal and clusters physical quantities.

\begin{figure}
\begin{center}
\includegraphics[width=0.32\textwidth,height=0.28\textwidth]{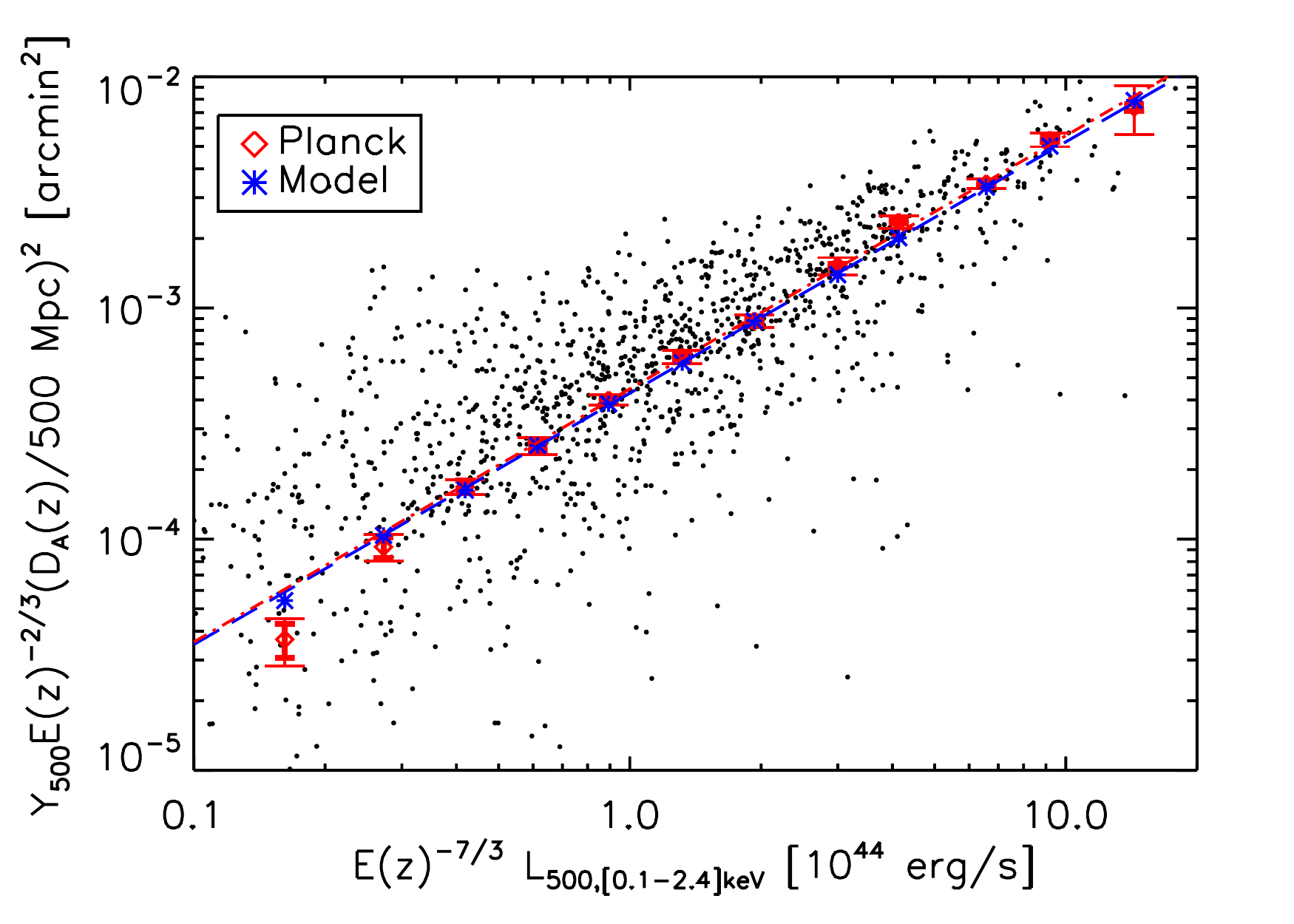}
\includegraphics[width=0.32\textwidth,height=0.275\textwidth]{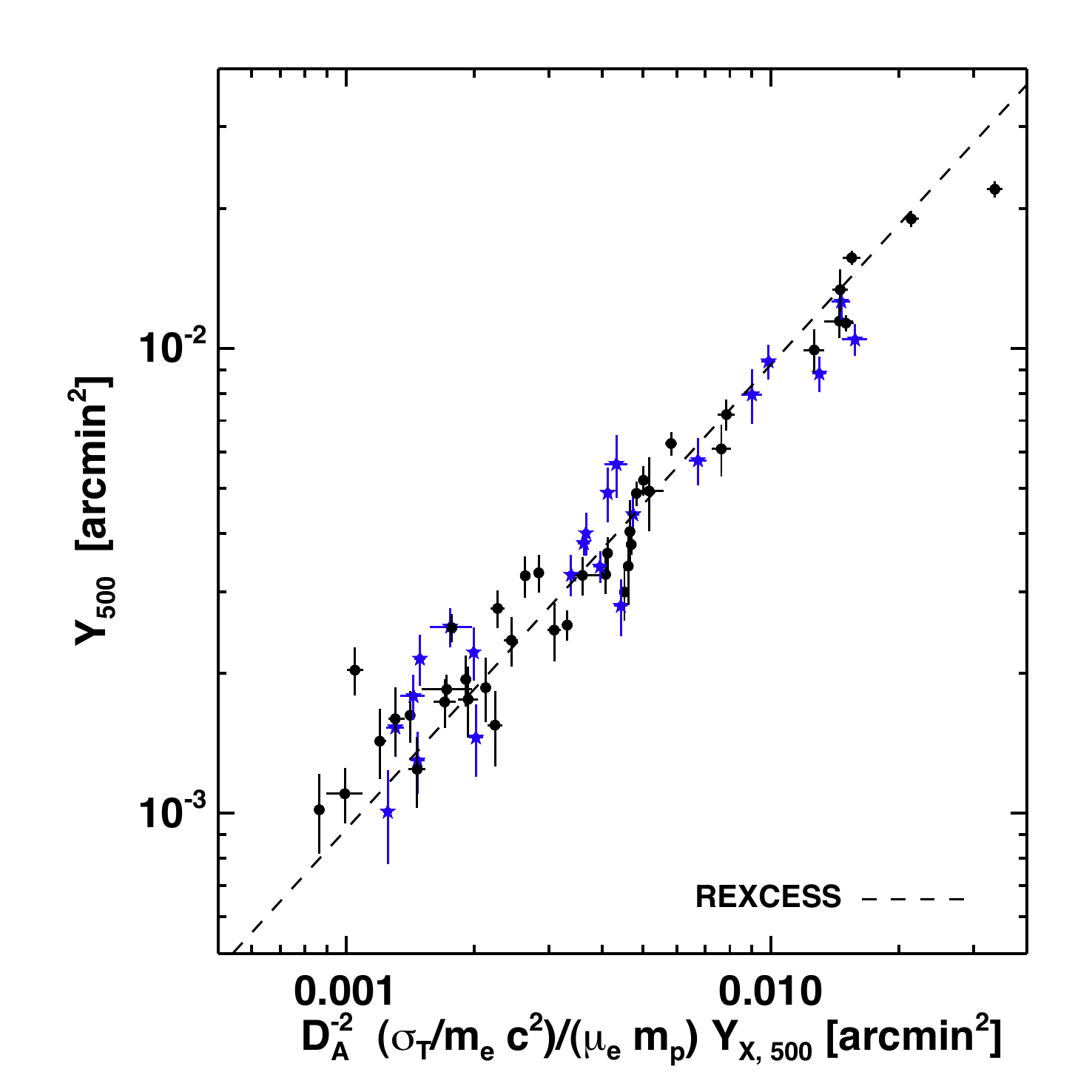}
\includegraphics[width=0.32\textwidth,height=0.26\textwidth]{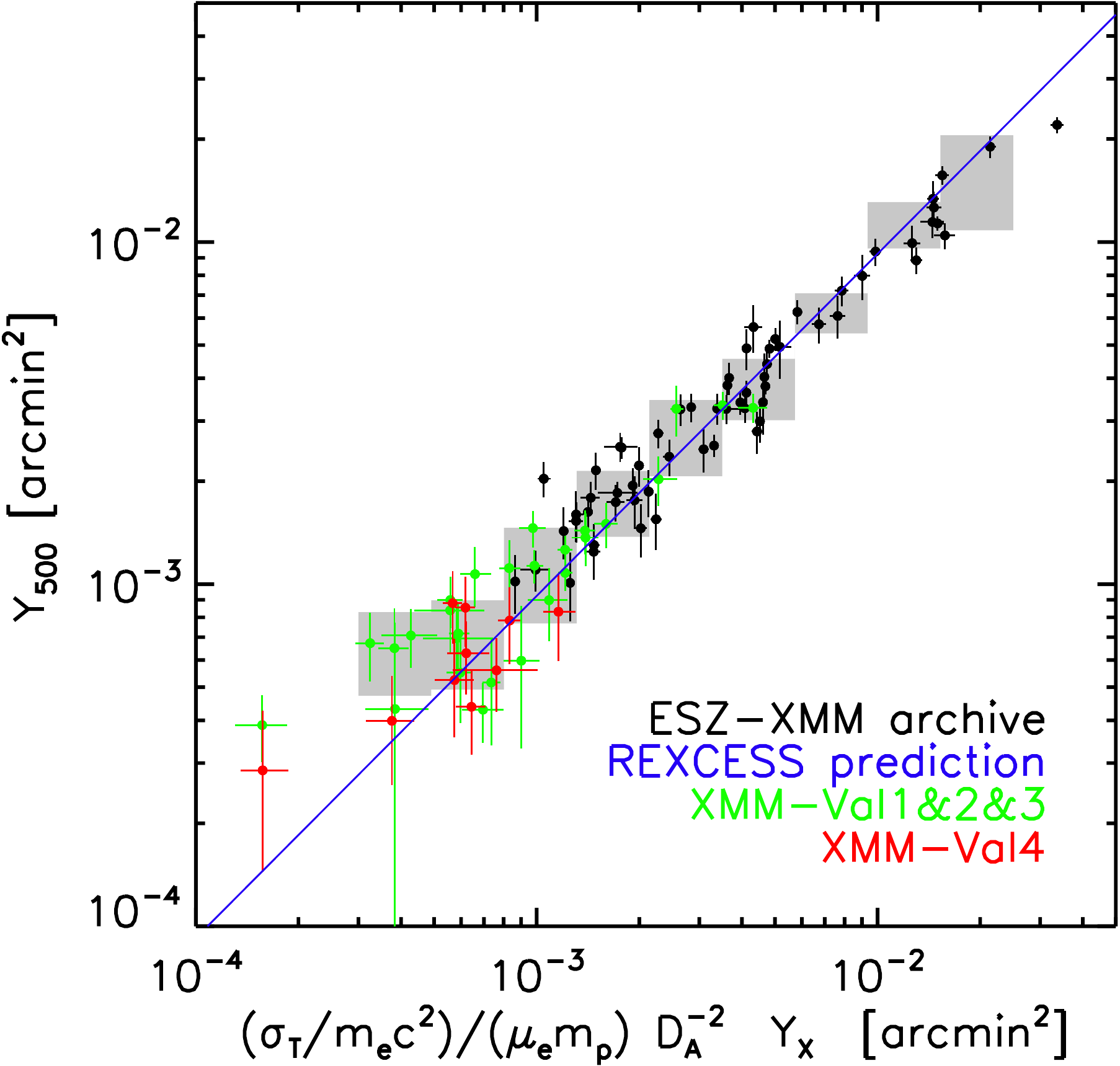}
\end{center}
\caption{{\bf Left panel:} 
Scaling relation between \Planck\ SZ measurements and X-ray luminosity for $\sim 1600$~MCXC clusters. Individual measurements are shown by the black dots and the corresponding bin averaged values by the red diamonds. Thick bars give the statistical errors, while the thin bars are bootstrap uncertainties. The X-ray based model is shown as a solid blue line, and the bin-averaged SZ cluster signal it predicts is shown by the blue stars. The red dot-dashed line shows the best fitting power-law to the data.
%\cite{planck2011-5.2a}\,. 
%
{\bf Middle panel:} SZ flux vs prediction from X-rays. Blue stars indicate cool core systems.  The dashed line is the prediction from \rexcess\ X-ray observations.
%\cite{arn10}\,. See [\refcite{planck2011-5.2b}]. 
%
{\bf Right panel:} Relation between apparent SZ signal ($\YSZ$) and the corresponding normalized $\YX$ parameter. Black points show clusters in the  \planck-ESZ sample with \xmm\ archival data;
% as presented in [\refcite{planck2011-5.2b}]; 
green and red points represent \Planck\ clusters confirmed with \xmm\ validation program. The blue lines denote 
the $\YSZ$ scaling relations predicted from the \rexcess\ X-ray observations.
%\cite{arn10}\,. 
The grey area corresponds to median $\YSZ$ values in $\YX$ bins with $\pm1\sigma$ standard deviation. 
}
\label{f:f3}
\end{figure}

The statistical combination of $\sim 1600$ MCXC clusters at $0.01<z<1$\,\cite{pif11} with the all-sky \planck\ data led to a precise measurement of the correlation between the SZ signal and the X-ray luminosity. Averaging SZ fluxes in bins of X--ray luminosity, $L_X$, we detected the SZ signal at very high significance. This  \planck\ observed signal is consistent with X-ray based predictions  over two decades in X--ray luminosity, down to  $L_X = 10^{43} {\rm erg/s} \lesssim L_{\rm 500} E(z)^{-7/3} \lesssim 2 \times 10^{45} {\rm erg/s}$. We found no deficiency in SZ flux with respect to the X--rays within $\Rv$. This results underlines the robustness and consistency of our overall view of intra-cluster medium properties (left panel of Fig.~\ref{f:f3}; reprinted from Fig. 4 in [\refcite{planck2011-5.2a}]). This analysis fully agrees with the similar study carried on beforehand on the WMAP-5 data by [\refcite{2012A&A...548A..51M}].
%by\cite{mel11}\,.

Moreover, it is also consistent with the more in-depth investigation of the local scaling relations conducted over a  sample of 62 massive known clusters detected by \planck\ at a high signal-to-noise ratio and with archival \xmm\ data\cite{planck2011-5.2b}\,. This analysis has allowed us to  investigate the scaling relations between the SZ signal, $D_{\rm A}^2\, Y_{500}$, and the X-ray-derived properties (i.e., gas mass $M_{\rm g,500}$, temperature $\TX$, luminosity $\LX$, SZ signal analogue $Y_{\rm X,500} = M_{\rm g,500} \times T_{\rm X}$, and total mass $\Mv$). The derived results are in excellent agreement with both X-ray predictions and recently-published ground-based data derived from smaller samples (middle panel of Fig.~\ref{f:f3}, reprinted from Fig. 4, left panel, in [\refcite{planck2011-5.2b}]; see [\refcite{and11,sifon12}]). 
  
Finally, as stressed in the previous section, the new clusters detected by \Planck\ follow the same scaling relations out to $z=1$  without significant deviation from self-similar evolution, exhibiting an equivalent agreement between their SZ and X-ray properties as show on the right panel of Fig.~\ref{f:f3} (reprinted from Fig. 7 in [\refcite{planck2012-IV}]; see [\refcite{planck2011-5.2a,planck2012-I}]). This behavior is seen down to an SZ signal of $\Yv\sim 3\times 10^{-4}$~arcmin$^2$. Below this threshold, we reach the current detection limit of \planck\ and Malmquist bias clearly appears (for details see [\refcite{planck2012-IV}]). 

%%%%%%%%%%%%%%%%%%%%%%%%%%%%
%%%%%%%%%%%%%%%%%%%%%%%%%%%%

As pointed out recently by  [\refcite{angulo12}], some observational and/or survey biases may arise from the combination of different effects and systematic biases. 
On the side of observable biased, 
one can mentioned the well known hydrostatic equilibrium hypothesis  which biases the X-ray masses low with respect to the true mass by 10-20\%\,\cite{kay04,piffaretti08}\,. The richness and weak lensing mass estimators from optical observations require a better control of their individual and statistical measurements\cite{johnston07,rozo09,biesiadzinski12,seh12,angulo12}\,.
Finally for SZ measurements, cross-calibration between \Planck, SPT and ACT measurement are certainly needed to further lower the photometric uncertainties and assess possible SZ flux measurement systematics.

Some of the aforementioned biases have been investigated on an individual cluster basis in [\refcite{planck2012-III}], where the relation between the \Planck\ SZ signal and the mass was studied using total masses derived from both weak lensing (WL) measurements\cite{okabe08,okabe10}  
%\cite[Subaru Telescope observations,][]{okabe08,okabe10} 
and from XÐray data assuming hydrostatic equilibrium (HE; \xmm\ observations). While the $M_{\rm WL} - D_{\rm M}^2 Y$ relation was consistent with previous measurements using WL 
%masses\cite{mar11}\,, 
masses\cite{Marriage11}\,, there was an offset in normalization with respect to the relation obtained using HE X-ray mass measurements. Since both the SZ measurements and the HE X-ray masses were consistent with our previous work, we concluded that the normalization offset in the $M_{\rm WL} - D_{\rm M}^2 Y$ is due to the X-ray masses being $\sim 20$ per cent higher than the WL masses. This is an unexpected result, given that simulations generally predict that HE X-ray masses should be smaller than WL masses owing to a the neglect of pressure support from bulk gas motions in the HE mass equation. Further investigation showed that the discrepancy is enhanced in dynamically disturbed systems and appears correlated with differences in mass concentration and the offset between the X-ray peak and the BCG position (the centers used for the X-ray and Wl mass determinations, respectively). More work is clearly needed, as discussed extensively in [\refcite{planck2012-III}].
These remaining 10-20\% inconstancies in scaling relations between SZ, X-ray and optical data are at hand. They need to be further investigated and quantified in order to reach an holistic view of the galaxy cluster properties.

Together with the SPT and ACT telescopes, the \Planck\ survey is shading new light on the population of galaxy clusters complementing our existing view of the ICM hot gas from the X-ray observations with high precision multi-frequency sub-millimeter to centimeter measurements. The scaling properties of the SZ signal together with other cluster observables have been investigated with various means and methods. 
Well constrained scaling relations between the SZ and X-ray measurements have been derived, with high precision calibration for the $\Yv-Y_{\rm X, 500}$, $\Yv-L_{\rm X, 500}$ and $\Yv-M$ relations. These results emphasize the well consistent picture we have of the ICM  at least within $\Rv$. 
Further \Planck\ intermediate results are currently being published providing further insights on the clusters of galaxies. 
%The nominal \Planck\ mission, which includes 15th months of data corresponding to two full sky surveys is to be released early 2013 to the community. 
%It will have a tremendous legacy value for the coming decades.

%%%%%%
% Neutrino mass from SZ surveys

\subsection{Neutrino mass from SZ surveys}

We explore here the possibility of setting useful constraints on the total neutrino 
mass from cluster number counts obtained by the ongoing {\it Planck}/SZ and future 
cosmic-variance-limited surveys. The precision with which this mass can be 
determined from SZ number counts is limited mostly by uncertainties in the 
cluster mass function and intra-cluster gas evolution. We find that projected 
{\it Planck}/SZ cluster counts could yield the total neutrino mass with a ($1\sigma$) 
uncertainty of $0.06$ eV, assuming the mass is in the range $0.1-0.3$ eV, and 
the survey detection limit is set at the $5\sigma$ significance level. Based on 
expected results from future cosmic-variance-limited (CVL) SZ survey, we predict 
a $1\sigma$ uncertainty of $0.04$ eV, a level comparable to that expected when 
CMB lensing extraction is carried out with the same experiment.

%\section{Summary}	

CMB measurements already placed meaningful upper limits on the total neutrino mass 
from its impact on the early integrated Sachs-Wolfe effect. The energy scale 
of recombination, $\sim 0.3$ eV, sets this upper limit; if the total 
neutrino mass is larger than this value, then neutrinos are non-relativistic 
and do not contribute to the decay of gravitational potentials shortly after 
recombination. If, on the other hand, the total mass is lower they constitute 
a relativistic component that contributes to the decay of linear gravitational 
potentials, changing the temperature of the CMB towards these gravitational wells.

Applying optimal estimators to CMB temperature and polarization maps one can recover 
the lensing potential to the precision that will allow constraining the total 
neutrino mass to the $0.04$ eV level (see [\refcite{2003PhRvL..91x1301K}]) 
%(Kaplinghat, Knox \& Song, 2003) 
with a cosmic-variance-limited (CVL) CMB experiment, assuming full-sky coverage, no 
foregrounds, and no source of non-Gaussianity other than the lensing of the CMB. 
In practice, it is unlikely that all these conditions will be fully satisfied and 
in that sense the frequently-quoted value $0.04$ eV is likely to be unrealistic.

Cluster number counts are yet another useful probe of neutrino masses. This 
is due to the fact that typical cluster scales are much smaller than the 
$\sim 150$ Mpc scale of linear dark matter halos that lens the CMB. In addition, 
cluster number counts are exponentially sensitive to $\sigma(M,z)$, the rms mass 
fluctuation on a cluster mass scale $M$ at redshift $z$, and since $\sigma(M,z)$ 
itself is exponentially sensitive to neutrino mass (via the growth function), this 
implies that cluster number counts should be a rather sensitive probe of neutrino 
masses
%, as has already been demonstrated in several previous works 
(see e.g. [\refcite{rephpap2,rephpap3}]).
 
We further explored the ability to strengthen the constraints on the neutrino mass 
from cluster number counts, and extended our forecast to additional surveys. This 
was done by parameterizing uncertainties in the halo mass function, which is 
the dominant source of modeling uncertainties. The shape and normalization of 
the mass function reflect the details of the growth of density fluctuations, and 
the nonlinear collapse and merger of sub-structures, whose hierarchical evolution 
can be best studied by state-of-the-art, large-volume hydrodynamical cosmological 
simulations. Currently available numerical codes predict a range of mass functions; 
this indeterminacy largely sets the precision limit of forecasting the total neutrino 
mass from cluster SZ number counts and power spectra. Additionally, we have accounted 
for cluster sample variance errors (in addition to Poissonian noise), a more realistic 
intra-cluster gas profile, as well as gas evolution with cluster mass and redshift. 
A full description of this work and results is given by ]\refcite{RephaeliShimonMG13}].

Our analysis shows that from cluster number counts alone (and priors based on measurements 
of the primary CMB APS and the HST prior on $H_{0}$), the uncertainty in 
the determination of the total neutrino mass can be limited to the $\sim 0.04-0.06$ 
eV range, depending on the details of the SZ cluster surveys and the fiducial neutrino 
mass. CMB anisotropy data combined with {\it Planck} cluster number counts are predicted 
to reach a level uncertainty at the higher end of this interval, whereas a CVL SZ 
survey is predicted to yield the somewhat higher precision corresponding to the 
lower end of this mass uncertainty interval. These results are based on the 
mass function by [\refcite{rephpap4}], whose parameter values were assumed to have uncertainties 
that are higher by 10\% than those specified there. 

\section{Selected topics in CMB studies}
\label{cmb}

The release of first cosmological products and papers from the {\it Planck} mission, waited for early 2013,  
will have a strong impact for cosmology in the coming decades, following the very important results from WMAP 
and recent ground-based projects together covering a wide multipole range. In this section  we discuss three very different topics: the first is connected to fundamental physics 
results expected in next times from the {\it Planck} mission; the second, regarding the polarization imprints induced by galaxy clusters and filaments, is relevant in particular for future high resolution 
ground-based experiments; the last concerns the information on primordial power spectrum at extremely high wavenumbers that could be derived from next generations 
of CMB spectrum missions, thus linking absolute measures of the CMB monopole, i.e. the largest angular scale, to small scale phenomena.

%%%%%%
% CMB anisotropy: Parity

%\newcommand{\be}{\begin{equation}}
%\newcommand{\ee}{\end{equation}}

\subsection{Fundamental physics from CMB Parity analyses}

The statistical properties of the CMB pattern may be used to constrain Parity (P) symmetry. 
Parity violations arise in several models: as modification of electromagnetism\cite{carroll90,carroll91,carroll97} or as modification of the standard picture of the Inflationary mechanism, where P is broken due to primordial (chiral) gravitational waves\cite{Lue:1998mq,Saito:2007kt,Sorbo:2011rz}\,. Both of these scenarios predict non null cross-correlations between gradient and curl modes and scalar and curl modes in the CMB polarization pattern. However, chiral gravity induces such correlations at the CMB last scattering surface whereas cosmological birefringence induces them by rotating the  polarization plane during the CMB photon journey from its last scattering to us\cite{Gluscevic:2010vv}\,. We focus here mainly on cosmic birefringence, reporting  findings from Gruppuso et al. 2011\cite{Gruppuso:2011ci}\,. In addition, we discuss the claimed P anomaly found at large angular scales in the anisotropy intensity spectrum of the WMAP data, first claimed by Kim and Naselsky in 2010\cite{Kim:2010gf,Kim:2010gd,Gruppuso:2010nd,Aluri:2011wv}\,.  The latter is dubbed a parity anomaly in view of an observed discrepancy (in power) among even and odd multipoles, which behave differently under P transformation. However, there is no sound theoretical framework that could explain such a mismatch. If the effect is indeed due to fundamental physics, its appearance at large angular scales naturally suggests the possibility that a P violating mechanism is involved during an early phase of the universe. Other explanations exist: for a more conservative approach see [\refcite{Kim:2010gd}] where it is conjectured that we may live in a special location of the universe, such that translational invariance is violated at scales larger than $\sim 4$ Gpc. 

\subsubsection{TT Parity anomaly}
\label{TTPanomaly}
All-sky temperature maps, $T(\hat n)$, are usually expanded in terms of spherical harmonics 
$Y_{\ell m}(\hat n)$, with $\hat n$ being a unit vector or direction on the sky,
completely specified by a couple of angles $(\theta, \phi)$. The quantities
$a_{T, \ell m} = \int d\Omega \, Y^{\star}_{\ell m}(\hat n) \, T(\hat n) \, ,$
are coefficients of the SH expansion, and $d \Omega = d \theta d \phi \sin \theta$. 
Under reflection (or P) symmetry ($\hat n \rightarrow -\hat n$), these coefficients behave as $ a_{T, \ell m} \rightarrow (-1)^{\ell} \,a_{T, \ell m}$.
CMB physics does not distinguish between even and odd multipoles\cite{Kim:2010gf,Kim:2010gd}\,. Therefore the power contained in even and odd multipoles must be statistically the same.  We thus define the quantity:
\be
C^{X}_{+/-} \equiv {1\over {(\ell_{max}-1)}} \, \sum_{\ell=2,\ell_{max}}^{+/-} {\ell (\ell + 1) \over{2 \pi }} \, \hat{C}^{X}_{\ell} 
\label{C+-}
\ee
where $\hat{C}^{X}_{\ell}$ are power spectral estimates for X = TT, TE, EE and BB.
The sum is meant only over the even or odd $\ell$ and this is represented respectively by the symbol $+$ or $-$. 
Therefore, two estimators can be built from Eq.~(\ref{C+-}): the "ratio" $ R^X = C^X_+/C^X_- \, $ (see [\refcite{Kim:2010gf,Kim:2010gd,Gruppuso:2010nd}]) and the 
"difference" $ D^X=C^X_+ - C^X_- \, $ (see [\refcite{Gruppuso:2010nd,Paci:2010wp}]), 
where $C^X_{\pm}$ is the band power average contained in the even (+) or odd (-) multipoles. 
\begin{figure}[t]
\centerline{
\includegraphics[angle=0,width=6cm]{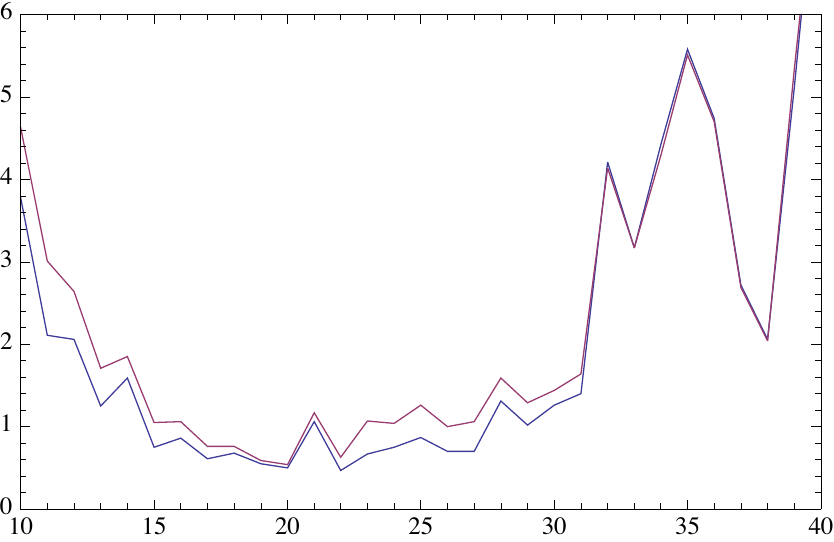}}
\caption{TT. Percentage of the WMAP 7 year value (y-axis) vs $\ell_{max}$ (x-axis). 
Blue line is for the ratio and the red line for the difference. 
%Reprinted from [\refcite{Gruppuso:2011ci}]. 
\label{parity_ell}}
%This behavior is consistent to what found by \cite{Kim:2010gd}.
\end{figure}
In Fig.~\ref{parity_ell} (reprinted from [\refcite{Gruppuso:2011ci}]) we plot 
%(from [\refcite{Gruppuso:2011ci}]) 
the percentage related to the WMAP 7 year P anomaly for TT versus $\ell_{max}$ in the range $10-40$ for the two considered estimators.
As evident there is not a single $\ell_{max}$ for which the TT anomaly shows up, but rather a characteristic scale in the $\ell$ range $[15,25]$.
We confirm the previously reported P anomaly in TT in the range $\Delta \ell=[2,22]$ at $> 99.5\%$ C.L..
{\it Planck} will not improve the signal-to-noise ratio in this range for the TT APS, since it is already cosmic variance dominated in the WMAP data.
However, {\it Planck} has a wider frequency coverage and this will improve the component separation layer in the data analysis pipeline. 
Moreover, {\it Planck} is observing the sky with a totally different scanning strategy and this represents a benefit for the analysis of systematic effects.

\subsubsection{Cosmological birefringence}
\label{PCMB}
Linear polarization maps  are components of a rank two tensor\cite{Zaldarriaga:1996xe} and are decomposed by the  spin harmonics $a_{\pm 2, \ell m} =  \int d\Omega \, Y^{\star}_{\pm 2, \ell m}(\hat n) \, ( Q(\hat n) \pm i U(\hat n))$, where $Y_{\pm 2, \ell m}(\hat n) $ are  SH of spin $2$ and $a_{\pm 2, \ell m}$ are the corresponding coefficients. It is then useful to introduce new coefficients as linear combinations of the previous: $a_{E, \ell m} = -(a_{2, \ell m} +a_{-2, \ell m} )/2$ and $a_{B, \ell m} = -(a_{2, \ell m} -a_{-2, \ell m} )/ 2 i$.
These have opposite behaviors under a P transformation: $a_{E, \ell m} \rightarrow (-1)^{\ell} \,a_{E, \ell m}$, $a_{B, \ell m} \rightarrow (-1)^{\ell+1} \,a_{B, \ell m}$. If P is conserved, by combining the previous transformation one immediately derives that the cross-correlations
$C_{\ell}^{TB} = \langle a_{T, \ell m}^{\star} a_{B, \ell 'm'}\rangle$ and $C_{\ell}^{EB} = \langle  a_{E, \ell m}^{\star} a_{B, \ell 'm'}\ \rangle$ must vanish. Further details can be found in 
[\refcite{Zaldarriaga:1996xe,Zaldarriaga:1997yt}] and explicit algebra is set forth in the Appendix 
of [\refcite{Gruppuso:2010nd}]. Parity violation could, however, may change this scenario. A popular model for which parity is broken in the photon sector is the Chern-Simons perturbation to the 
Maxwell Lagrangian\cite{carroll90}\,:  $ \Delta{\mathcal L}= -\frac{1}{4}\,  p_{\mu} \epsilon^{\mu\nu\rho\sigma}F_{\rho\sigma} A_{\nu}$,  where ${F}^{\mu\nu}$ is the Maxwell tensor and $A^\mu$ the four-potential. One of the consequences is in vacuo dispersion of photons, in particular those from the CMB and the rotation of their polarization planes, observable through $TB$ and $EB$ correlations, that acquire a signal modulated by $\alpha$ (or ``rotated'')\cite{Lue:1998mq,Feng:2006dp,Feng:2004mq,Komatsu:2008hk,Cabella:2007br}\,.

The WMAP team\cite{Komatsu:2010fb} reported $\alpha^{{\rm WMAP} \, 7 yr} = -0.9^{\circ} \pm 1.4^{\circ}$ at $68\%$ C.L.. 
Our constraint, obtained at low resolution \cite{Gruppuso:2011ci} and considering the same estimator that has been used in [\refcite{Wu:2008qb}], 
reads $\alpha = -1.6^{\circ} \pm 1.7^{\circ} \, (3.4^{\circ})$ at $68\%$ ($95\%$) C.L. for $\Delta \ell = 2-47$.
Considering $\Delta \ell = 2-23$ we obtain $\alpha = -3.0^{\circ +2.6^{\circ}}_{\phantom{a}-2.5^{\circ}}$ at $68\%$ C.L. and 
$\alpha = -3.0^{\circ +6.9^{\circ}}_{\phantom{a}-4.7^{\circ}}$ at $95\%$ C.L.. This is the same multipole range considered by the WMAP 
team at low resolution in [\refcite{Komatsu:2010fb}] (the only other result available in the literature at these large angular scales) 
where  with a pixel based likelihood analysis they obtain $\alpha^{{\rm WMAP} \, 7 yr} = -3.8^{\circ} \pm 5.2^{\circ}$ at $68\%$ C.L..
In [\refcite{Gubitosi:2009eu}] it is claimed that the improvement expected for the {\it Planck} 
%satellite\cite{Ade:2011ah} 
satellite\cite{planck2011-1.1} 
in terms of sensitivity\cite{blubook} is 
around $15$. Almost the same number is obtained in Gruppuso et al. [\refcite{Gruppuso:2011ci}]. Both  forecasts are provided considering just 
the nominal sensitivity whereas the uncertainties coming from the systematic effects are not taken into account.

%%%%%%
% Clusters, fliaments and CMB Polarization

\subsection{CMB induced polarization from single scattering by clusters of galaxies and filaments}

We discuss here two types of secondary polarization 
effects arising from single scattering of the CMB photons by ionized gas. 
These are the CMB quadrupole induced polarization (pqiCMB), which couples 
the gas density with the CMB quadrupole component, and the polarization 
induced by the gas motion transverse to the line of sight
(p$\beta_t^2$SZ). 
%Both effects are directly proportional to the optical depth. 

\begin{figure}[t]
\centerline{\psfig{file=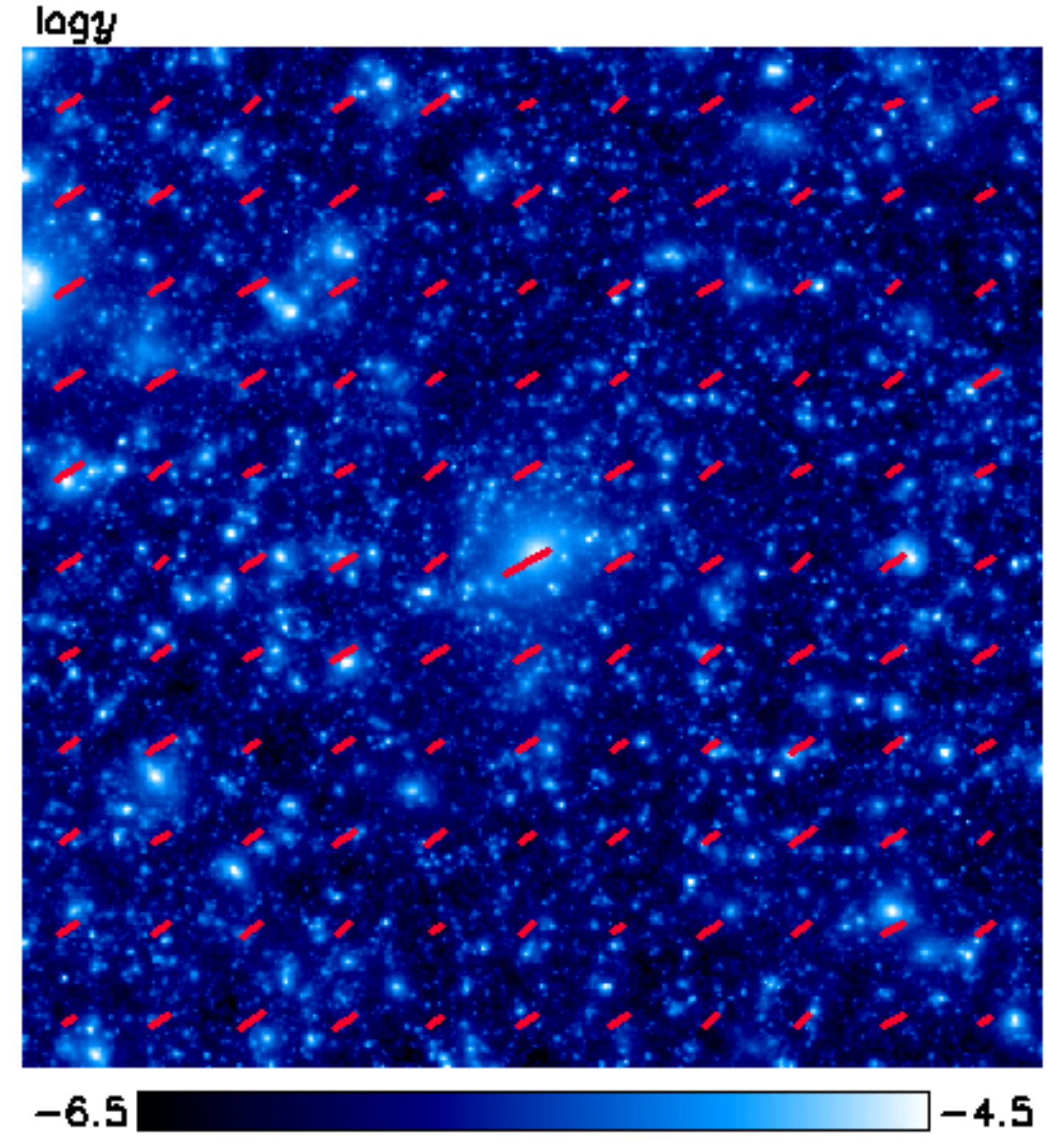,width=4.cm}\psfig{file=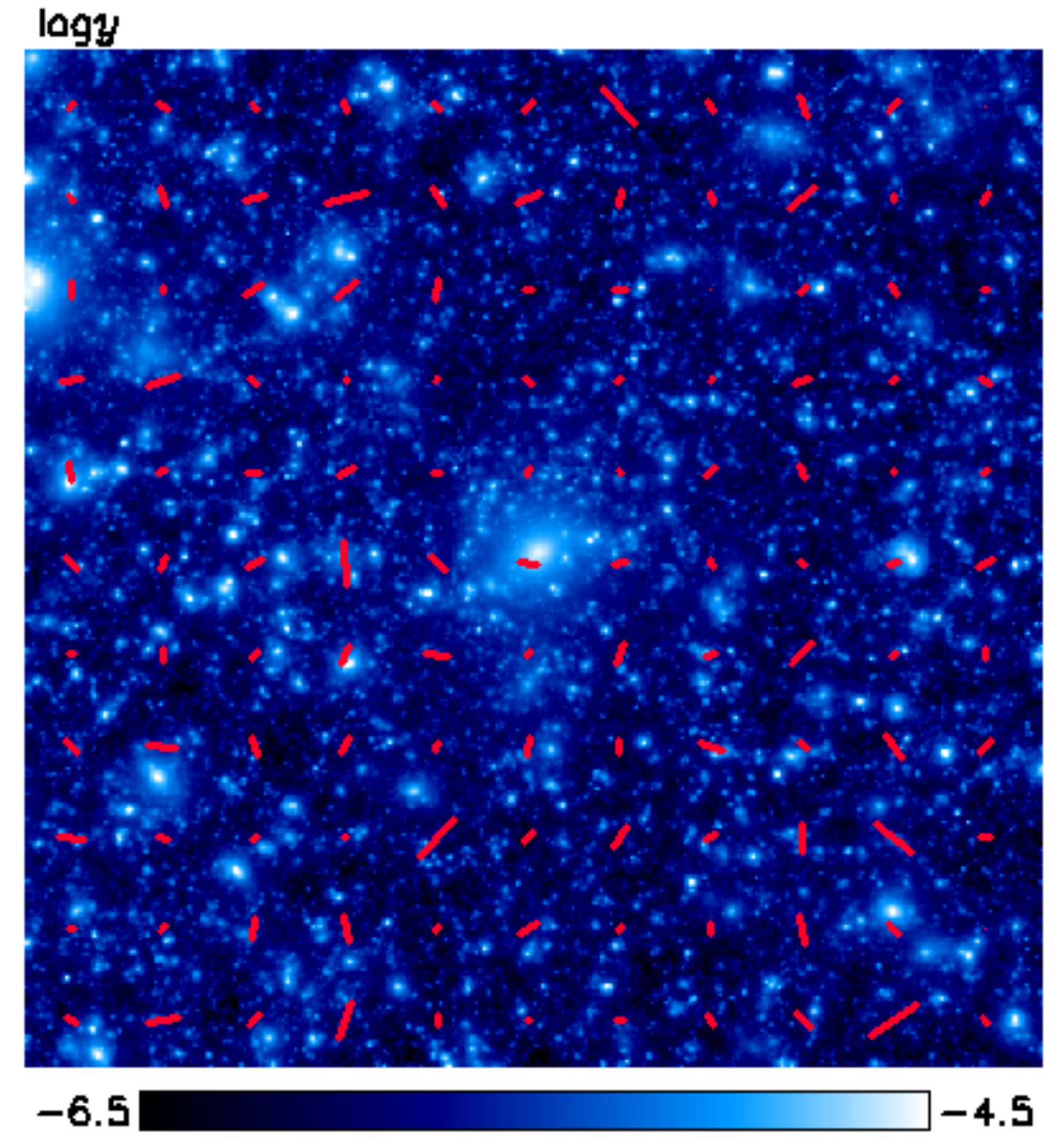,width=4.cm}}
\vspace*{8pt}
\caption{Polarization map (red vectors) of
  0.93$^\circ\times$0.93$^\circ$ integrated in redshift of one sky
  patch realization for pqiCMB (left panel) and p$\beta_t^2$SZ
(right panel). For comparison, we put the logarithm of Compton $y$-parameter (color scaled). 
\label{fig_liu}}
\end{figure}

Instead of using an analytical model adopted in many previous studies
or simulations of individual galaxy clusters (see e.g. [\refcite{liu,corybau03,lavx04,shi06}]), we used
high-resolution N-body/Hydrodynamic simulations featuring adiabatic gas physics and a
novel box-stacking  scheme that allows to reconstruct the CMB
quadruple component and  the physical properties of the scattering
media along the light cone traversed by radiation. We generated 28
random sky patches integrated along the light-cone, each of about 0.86
deg$^{2}$ at angular resolution of 6''. The primordial CMB quadrupole information in
each simulation box is computed by inverse Fourier transform of CMB
quadrupole components in Fourier space at
all conformal times required by the map-making strategy. For each
individual Fourier mode, we decide the initial value $\Psi_i({\bf k})$
by drawing a random number from a Gaussian distribution with variance
given by the initial power spectrum obeys a power law,
$P_{\Psi}(k)=Ak^{n_s-4}$  with $A$ a normalization factor and $n_s$ a
spectral index of the scalar perturbations. The time evolution of CMB
quadrupole in each individual mode is computed using the
CMBFast\cite{CMBFast} Boltzmann code (see [\refcite{Ramos}] for more details).
Here we focus on the characterization of the polarized signals in the
simulations  and the study of their statistical properties at high angular
resolution.  

The results from the pixel distribution in the frequency independent
maps, show that the linear polarization degrees follows,
in logarithm scale, nearly Gaussian distributions, centered
around $10^{-8}$ and $10^{-10}$ for   
pqiCMB and p$\beta_t^2$SZ, respectively. Our simulations confirm that
the polarization degree of the pqiCMB is  
a close proxy of the electron density column and that the polarization 
angles of this effect are closely aligned due to the slow variation of 
primary CMB quadrupole on small sky patches and along the line of
sight (see Fig.~\ref{fig_liu}, reprinted from [\refcite{Ramos}], left panel).
The effect of the gas overlap along
density columns causes galaxy clusters and other bound objects
to be less prominent (with respect to the mean background) than
in the case of the thermal SZ effect, where the signal in clusters
is boosted by the high temperature. In the case of  p$\beta_t^2$SZ
(right panel of  Fig.~\ref{fig_liu}), the
polarization degrees and angles are weighted by the transverse
velocity of the scattering media, therefore the integration along
the line of sight can erase contributions from individual collapsed
objects, depending on their internal velocity structure and the
effect of gas overlap.

By producing maps of these secondary induced polarization effects at
different frequencies, we confirm the strong dependence on frequency
of both signals, especially in the case of p$\beta_t^2$SZ, for 
which the mean value increases by a factor of $\sim 100$ from
the 30 GHz to 675 GHz. The high magnitudes of both signals at
high frequencies may allow its detection with the next generation
of sub-millimeter instruments. 

The redshift distribution of the polarization degrees shows
that the contribution for the polarization signal is highest at $z \simeq 1$ and $z \simeq 0.5$ for the pqiCMB and p$\beta_t^2$SZ,
respectively. %We also note that 
Finally, only about 7\% of the total signal comes from $z > 4$
for the former and $z > 3$ for the latter and both signals converge
rapidly at larger $z$.
%above these redshifts.

%%%%%%
% CMB Spectrum

\def\aap{A\&A}
\def\apj{ApJ}
\def\apjs{ApJS}
\def\apjl{ApJL}
\def\mnras{MNRAS}
\def\aj{AJ}
\def\nat{Nature}
\def\aaps{A\&A Supp.}
\def\pra{Phys.Rev.A}         % Physical Review A: General Physics
\def\physrep{Physics Reports}         % Physical Review A: General Physics
\def\prb{Phys.Rev.B}         % Physical Review B: Solid State
\def\prc{Phys.Rev.C}         % Physical Review C
\def\prd{Phys.Rev.D}         % Physical Review D
\def\prl{Phys.Rev.Lett}      % Physical Review Letters
\def\araa{ARA\&A}       % Annual Review of Astron and Astrophys
\def\gca{GeCoA}         % Geochimica et Cosmochimica Acta
\def\pasp{PASP}              % Publications of the ASP
\def\pasj{PASJ}              % Publications of the ASJ
\def\apss{ApSS}
\def\jcap{JCAP}
\def\sovast{Soviet Astronomy}
\def\na{New Astronomy}
\newcommand{\DTe}{{{\Delta_{T_e}}}}
\newcommand{\SQ}{{{\mathcal{E}}}}
\newcommand{\TBB}{{{T_{\rm BB}}}}
\newcommand{\rs}{{{r_{\rm s}}}}
\newcommand{\TBE}{{{T_{\rm BE}}}}
\newcommand{\TCMB}{{{T_{\rm CMB}}}}
\newcommand{\Te}{{{T_{\rm e}}}}
\newcommand{\Teq}{{{T^{\rm eq}_{\rm e}}}}
\newcommand{\Ti}{{{T_{\rm i}}}}
\newcommand{\nB}{{{n_{\rm B}}}}
\newcommand{\nHe}{{{n_{\rm ^4He}}}}
\newcommand{\nH}{{{n_{\rm H}}}}
\newcommand{\nHet}{{{n_{\rm ^3He}}}}
\newcommand{\nHt}{{{n_{\rm { }^3H}}}}
\newcommand{\nHtw}{{{n_{\rm { }^2H}}}}
\newcommand{\nBes}{{{n_{\rm { }^7Be}}}}
\newcommand{\nBE}{{{n_{\rm BE}}}}
\newcommand{\Bes}{{{{\rm { }^7Be}}}}
\newcommand{\nLis}{{{n_{\rm { }^7Li}}}}
\newcommand{\nLisi}{{{n_{\rm { }^6Li}}}}
\newcommand{\nS}{{{n_{\rm S}}}}
\newcommand{\nSS}{{{n_{\rm ss}}}}
\newcommand{\Teff}{{{T_{\rm eff}}}}
\newcommand{\Mpc}{{{{\rm ~Mpc}}}}
\newcommand{\id}{{{\rm d}}}
\newcommand{\aR}{{{a_{\rm R}}}}
\newcommand{\bR}{{{b_{\rm R}}}}
\newcommand{\neb}{{{n_{\rm eb}}}}
\newcommand{\neql}{{{n_{\rm eq}}}}
\newcommand{\kB}{{{k_{\rm B}}}}
\newcommand{\EB}{{{E_{\rm B}}}}
\newcommand{\zmin}{{{z_{\rm min}}}}
\newcommand{\zmax}{{{z_{\rm max}}}}
\newcommand{\zinj}{{{z_{\rm inj}}}}
\newcommand{\YBEC}{{{Y_{\rm BEC}}}}
\newcommand{\yg}{{{y_{\rm \gamma}}}}
\newcommand{\y}{{{y}}}
\newcommand{\rhob}{{{\rho_{\rm b}}}}
\newcommand{\Ne}{{{n_{\rm e}}}}
\newcommand{\sigT}{{{\sigma_{\rm T}}}}
\newcommand{\me}{{{m_{\rm e}}}}
\newcommand{\npl}{{{n_{\rm pl}}}}
\newcommand{\nY}{{{n_{\rm Y}}}}
\newcommand{\kD}{{{{k_{\rm D}}}}}
\newcommand{\KC}{{{{K_{\rm C}}}}}
\newcommand{\KdC}{{{{K_{\rm dC}}}}}
\newcommand{\Kbr}{{{{K_{\rm br}}}}}
\newcommand{\zdC}{{{{z_{\rm dC}}}}}
\newcommand{\zbr}{{{{z_{\rm br}}}}}
\newcommand{\aC}{{{{a_{\rm C}}}}}
\newcommand{\adC}{{{{a_{\rm dC}}}}}
\newcommand{\abr}{{{{a_{\rm br}}}}}
\newcommand{\gdC}{{{{g_{\rm dC}}}}}
\newcommand{\gbr}{{{{g_{\rm br}}}}}
\newcommand{\gff}{{{{g_{\rm ff}}}}}
\newcommand{\xe}{{{{x_{\rm e}}}}}
\newcommand{\alphafs}{{{{\alpha_{\rm fs}}}}}
\newcommand{\YHe}{{{{Y_{\rm He}}}}}
\newcommand{\SE}{{{\dot{{Q}}}}}
\newcommand{\SN}{{\dot{{N}}}}
\newcommand{\muc}{{{{\mu_{\rm c}}}}}
\newcommand{\xc}{{{{x_{\rm c}}}}}
\newcommand{\xH}{{{{x_{\rm H}}}}}
\newcommand{\mT}{{{{\mathcal{T}}}}}
\newcommand{\mG}{{{{\mathcal{G}}}}}
\newcommand{\Ob}{{{{\Omega_{\rm b}}}}}
\newcommand{\Or}{{{{\Omega_{\rm r}}}}}
\newcommand{\Odm}{{{{\Omega_{\rm dm}}}}}
\newcommand{\mdm}{{{{m_{\rm WIMP}}}}}

\subsection{Mixing of blackbodies: creation of entropy and dissipation of
  sound waves in the early Universe}

There is a very important connection between the spectrum of the monopole or sky
averaged CMB, which is an almost perfect blackbody and  COBE/FIRAS\cite{cobe} detected no deviation from Planck spectrum, and
the angular anisotropies precisely measured by WMAP \cite{wmap7}\,,
SPT\cite{spt}\,, ACT\cite{act}\,, {\it Planck} and other experiments on scales corresponding to
comoving wavenumber $10^{-4}\lesssim k \lesssim 0.2~\rm{Mpc}^{-1}$,
including the damping tail due to photon diffusion\cite{silk,kaiser}\,. The power that disappears from the CMB APS 
because of Silk damping appears in the energy spectrum of monopole as $y$ \cite{zs1969}\,, $\mu$ and
intermediate-type  distortions\cite{bdd95,ks2012b}\,.   
The primordial power spectrum, at comoving
wavenumbers  $8\lesssim k \lesssim 10^4 ~\rm{Mpc}^{-1}$ (mostly inaccessible by any
other means), can thus be recovered by
precise measurements of the energy spectrum of the monopole.

\begin{figure}
\resizebox{\hsize}{!}{\includegraphics{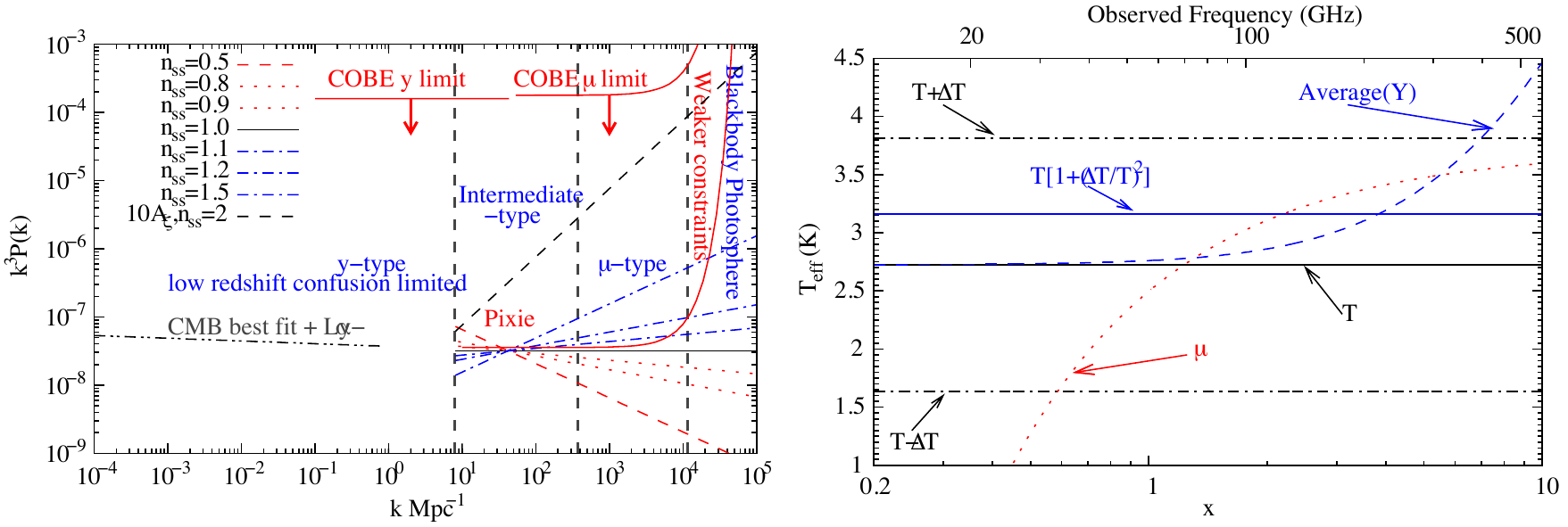}}
\caption{\label{bbmixfig} {\bf Left panel:} current and future constraints on
  primordial power spectrum as a function of comoving wavenumber
  $k$. Possible spectra on small scales allowed by current data are also
  shown. {\bf Right panel:} the spectrum resulting from mixing of blackbodies. We have used the linear
(in $(\Delta T/T)^2$)
solution to make the plots but used a large value of $\Delta T/T$ to make
the differences visible. 
Effective temperature of the spectrum
defined by writing the occupation number as $n=1/(e^{h\nu/(\kB\Teff)}-1)$ as a function of
dimensionless frequency, $x=h\nu/kT$ is plotted. 
%Figures are taken from [\refcite{ks2012b}] and [\refcite{ksc2012b}]. 
At high redshifts, $z\gtrsim 10^5$, the spectrum comptonizes
rapidly to create a $\mu$-type distortion or Bose-Einstein spectrum, also
shown in the figure.}
\end{figure}

The current
constraints on the primordial power spectrum, including Ly-$\alpha$ forest constraints \cite{ly2006,ssm2006}\,, are shown in the left panel of
Fig.~\ref{bbmixfig} (reprinted from [\refcite{ks2012b}]). At present the small-scale constraints
from  COBE/FIRAS  $y$-type
($2\sigma$ limit $y\lesssim 1.5\times 10^{-5}$) and $\mu$-type ($\mu\lesssim 9\times
10^{-4}$) distortions are very weak and considerable freedom is allowed on
small scales. Proposed future experiment PIXIE\cite{pixie} would improve the
small-scale constraints by a factor of $\sim 2500$ and start probing the
interesting region of the parameter space, extending our knowledge of the
primordial power spectrum by many orders of magnitude in terms of the scales probed.

 Previous calculations
of  spectral distortions in CMB from  Silk damping \cite{sz1970b,daly1991,hss94}
underestimated the energy in sound waves and also assumed that all the
dissipated energy goes into creating spectral distortions.  The physics of creation of spectral
distortion becomes very simple if we consider the fact that diffusion of
photons, which damps the CMB perturbations, is in fact mixing blackbodies
of different temperature \cite{cks2012,ksc2012b}\,. The right panel in Fig.~\ref{bbmixfig} 
(reprinted from [\refcite{ksc2012b}]) shows the
result of averaging two blackbodies with temperatures $T\pm \Delta T$.   The
resulting spectrum is marked 'Average(Y)' and is a $y$-type distortion
\cite{zis1972} on top of a blackbody with temperature
$T\left[1+\left(\Delta T/T\right)^2\right]$ with the two curves crossing at
$x=3.83$. The averaging of two blackbodies adds energy as well as photons
to the average CMB monopole and therefore not all the energy can be used to create
spectral distortions. It is straightforward to show, by using Taylor series
expansion of the initial blackbodies up to second order in $\Delta T/T$ and
then doing the ensemble average, that only $1/3$ of the dissipated energy goes into $y/\mu$-type
distortions \cite{ksc2012b} and $2/3$ just raises the temperature.  Applying the above procedure to CMB  immediately gives
us the rate of
energy injection into CMB and the resulting $\mu$ distortion \cite{is1975b}\,,
\begin{align}
\left.\frac{\id \mu}{\id t}\right|_{\rm distortion}=1.4\left.\frac{\id}{\id t}\frac{\Delta E}{E_{\gamma}}\right|_{\rm distortion}&=-1.4\frac{\id}{\id t}\frac{1}{3}6\int \frac{k^2\id k}{2\pi^2}P_i(k)\left[\sum_{\ell=0}^{\infty}(2\ell+1)\Theta
_{\ell}^2(k)\right]\nonumber\\
&\approx -\frac{\id}{\id t}2.8\int \frac{k^2\id k}{2\pi^2}P_i(k)\left[\Theta
_{0}^2+3\Theta_1^2\right],
\end{align}
where $\frac{\Delta E}{E_{\gamma}}$ is the fractional energy going into CMB
  distortion, $\Theta_{\ell}$ are the multipole moments of CMB temperature
  perturbation transfer function, $P_i$ is the initial power spectrum, and in the second line we have used the fact that during
  tight coupling the $\ell\ge 2$ modes are suppressed. The time derivatives
  are easily calculated using the tight coupling solutions with Silk damping
  or by using the first order Boltzmann equation. A nice
  feature of the approach presented above is that the energy injected into the distortion
  can be directly identified with the increase in entropy of CMB\cite{ksc2012b}\,.

%
%\section{References}
%  \bibliographystyle{ws-ijmpd}
%  \bibliography{CBsess_MG13review_Refs}
%

\bigskip
\bigskip

{\small
\noindent
{\bf Acknowledgements --}  We thank the Editorial Board of Astronomy and Astrophysics (European Southern Observatory; ESO) 
and the Editors of the Annual meeting of the French Society of Astronomy and Astrophysics
for having granted us the permission to reproduce many figures originally published (or in press) in the same Journals or Proceedings.
Some figures are reproduced by permission of the AAS (American Astronomical Society).
Credits are indicated when each reprinted figure is mentioned in the text for the first time. 
The authors that are members {\it Planck} Collaboration warmly thank the {\it Planck} Collaboration and, in particular, all the members of the {\it Planck}  Working
Groups 2, 5, 6, 7 and of the HFI and LFI Core Teams, with whom they shared the analysis and the interpretation of {\it Planck}  data as for the subjects
discussed here, and the members of the {\it Planck}  Science Team and Editorial Board for the permission of publishing this paper.
CB and PN wish to thank A. Gruppuso and N. Mandolesi, with whom have carried out part of the work described in this paper.
Some of the results in this paper have been derived using the HEALPIX\cite{gorski} package.
We acknowledge the use of the Legacy Archive for Microwave Background Data Analysis (LAMBDA) supported by the NASA Office of Space Science. 
Some of the simulations presented in this work have been performed using the computational facility of IASF Bologna and at CINECA.
CB, MM, AM, PN acknowledge support by ASI through ASI/INAF Agreement I/072/09/0 for the Planck LFI Activity of Phase E2 and by MIUR through PRIN 2009 grant n. 2009XZ54H2.
EP acknowledges the support of ANR project Multiverse under grant ANR-11-BD56-015. 
LT acknowledges partial financial support from the Spanish Ministerio de Econom\'\i{a} y Competitividad, under proyects AYA2010-21766-C03-01,
AYA2012-39475-C02-01 and by the Consolider Ingenio-2010 Programme, project CSD2010-00064.
}

%OTHER ACKNOWLEDGEMENTS ???

%\smallskip

%REFERENCES: TO BE ORDERED 

%\section*{References}

%\begin{thebibliography}{000} %for 3 digits
%\begin{thebibliography}{00}  %for 2 digits
%\begin{thebibliography}{0}    %for 1 digit

\end{document}